\documentclass[usenatbib]{mnras}
\usepackage{amsmath,amssymb}
\usepackage[pdftex]{graphicx}

\def\vb#1{\mbox{\boldmath $#1$}}

\newcommand{\Kel}{$\rm SMSS \ J0313-6708$}

\newcommand{\Caf}{$\rm SDSS \ J1029+1729$}

\newcommand{\percc}{{\rm cm^{-3}}}
\newcommand{\um}{{\rm \mu m}}
\newcommand{\E}[1]{\times 10^{#1}}

\newcommand{\nH}{n_{{\rm H}}}

\newcommand{\nHmax}{n_{\rm H,max}}

\newcommand{\HII}{H~{\sc ii}}

\newcommand{\Mhalo}{M_{\rm halo}}
\newcommand{\Mhalodm}{M_{\rm halo, dm}}
\newcommand{\Mhalob}{M_{\rm halo, b}}
\newcommand{\Rhalo}{R_{\rm halo}}

\newcommand{\ljeans}{\lambda _{\rm J}}
\newcommand{\Mjeans}{M_{\rm J}}

\newcommand{\MPopIII}{M_{\rm Pop III}}
\newcommand{\Esn}{E_{{\rm SN}}}

\newcommand{\tff}{t_{\rm ff}}

\newcommand{\tlife}{t_{\rm life}}
\newcommand{\tsn}{t_{\rm SN}}
\newcommand{\tret}{t_{\rm ret}}
\newcommand{\zret}{z_{\rm ret}}

\newcommand{\Mmet}{M_{\rm met}}

\newcommand{\QH}{Q({\rm H})}

\newcommand{\kmpers}{{\rm km \ s^{-1}}}

\newcommand{\cs}{c_{\rm s}}

\newcommand{\ffb}{f_{\rm ret}}

\newcommand{\abM}[1]{M_{\rm {#1}}}
\newcommand{\abAsun}[1]{A_{\bigodot}({\rm {#1}})}
\newcommand{\abAcr}[1]{A_{\rm cr}({\rm {#1}})}
\newcommand{\abA}[1]{A({\rm {#1}})}
\newcommand{\abH}[1]{{\rm [{#1}/H]}}

\newcommand{\abX}[1]{X({\rm {#1}})}
\newcommand{\abFe}[1]{{\rm [{#1}/Fe]}}
\newcommand{\muX}[1]{\mu_{\rm {#1}}}

\newcommand{\rhoamb}{\rho _{{\rm amb}}}

\newcommand{\XH}{X_{{\rm H}}}
\newcommand{\Zsun}{{\rm Z_{\bigodot}}}
\newcommand{\Msun}{{\rm M_{\bigodot}}}
\newcommand{\Msunyr}{\Msun/{\rm yr}}
\newcommand{\Mcut}{M_{\rm cut}}

\newcommand{\Dcrit}{{\cal D}_{{\rm cr}}}

\newcommand{\Mpr}{M_{{\rm pr}}}

\newcommand{\Rs}{R_{\rm sh}}

\defcitealias{Chiaki15}{C15}
\defcitealias{Chiaki16}{C16}
\defcitealias{Chiaki17}{C17}
\defcitealias{Chiaki18}{C18}
\defcitealias{Chiaki19}{Paper I}
\defcitealias{Smith15}{S15}
\defcitealias{Hirano14}{H14}
\defcitealias{Ishigaki14}{I14}
\defcitealias{Tominaga14}{T14}
\defcitealias{Marassi14}{M14}
\defcitealias{Marassi15}{M15}

\usepackage{color}
\definecolor{rev}{rgb}{0.8,0.0,0.0}

\title[CEMP star formation]
      {Seeding the second star --- II. CEMP star formation enriched from faint supernovae}

\author[G. Chiaki et al.]
{Gen Chiaki$^{1}$\thanks{E-mail: gen.chiaki@physics.gatech.edu},
John H. Wise$^{1}$,
Stefania Marassi$^{2}$,
Raffaella Schneider$^{2,3,4}$,
\newauthor
Marco Limongi$^{5,6,7}$, and
Alessandro Chieffi$^{7,8,9}$
\\
$^{1}$Center for Relativistic Astrophysics, School of Physics, Georgia Institute of Technology, Atlanta, GA 30332, USA \\
$^{2}$Dipartimento di Fisica, ``Sapienza'' Universit\`{a}di Roma, Piazzale Aldo Moro 5, 00185 Roma, Italy \\
$^{3}$INAF/Osservatorio Astronomico di Roma, Via di Frascati 33, 00040 Monte Porzio Catone, Italy \\
$^{4}$INFN, Sezione Roma1, Dipartimento di Fisica, ``Sapienza'' Universit\`{a}di Roma, Piazzale Aldo Moro 5, 00185, Roma, Italy \\
$^{5}$Istituto Nazionale di Astrofisica - Osservatorio Astronomico di Roma, Via Frascati 33, I-00040, Monteporzio Catone, Italy \\
$^{6}$Kavli Institute for the Physics and Mathematics of the Universe, Todai Institutes for Advanced Study, the University \\
of Tokyo,Kashiwa, Japan 277-8583 (Kavli IPMU, WPI) \\
$^{7}$INFN - Sezione di Perugia, via A. Pascoli, Perugia, Italy \\
$^{8}$Istituto di Astrofisica e Planetologia Spaziali, INAF, via Fosso del cavaliere 100, 00133 Roma, Italy \\
$^{9}$Monash Centre for Astrophysics (MoCA), School of Mathematical Sciences, Monash University, Victoria 3800, Australia \\
}

\begin{document}

\date{}

\pagerange{\pageref{firstpage}--\pageref{lastpage}} \pubyear{2020}

\maketitle

\label{firstpage}

\begin{abstract}
Carbon-enhanced metal-poor (CEMP) stars are the living fossils holding records of chemical enrichment
from early generations of stars.
In this work, we perform a set of numerical simulations of 
the enrichment from a supernova (SN) of a first generation of metal-free (Pop III) star and
the gravitational collapse of the enriched cloud, considering all relevant cooling/heating processes 
and chemical reactions as well as the growth of dust grains.
We adopt faint SN models for the first time with progenitor masses $\MPopIII = 13$--$80 \ \Msun$, 
which yield C-enhanced abundance patterns ($\abFe{C} = 4.57$--$4.75$)
through mixing and fallback of innermost layers of the ejecta.
This model also considers the formation and destruction of dust grains.
We find that the metals ejected by the SN can be partly re-accreted by the same dark matter minihalo,
and carbon abundance of the enriched cloud $\abA{C} = 3.80$--$5.06$ is
lower than the abundance range of observed CEMP stars ($\abA{C} \gtrsim 6$)
because the mass of the metals ejected by faint SNe is smaller than normal core-collapse SNe
due to extensive fallback.
We also find that cloud fragmentation is induced by gas cooling from carbonaceous grains 
for $\MPopIII = 13 \ \Msun$ even with the lowest iron abundance $\abH{Fe} \sim -9$.
This leads to the formation of low-mass stars, and these ``giga metal-poor'' stars can survive until
the present-day Universe and may be found by future observations.
\end{abstract}

\begin{keywords} 
  galaxies: evolution ---
  ISM: abundances --- 
  stars: formation --- 
  stars: low-mass --- 
  stars: Population III ---
  stars: Population II
\end{keywords}


\section{INTRODUCTION}

Metal-poor stars are important astronomical objects that 
hold the signatures of chemical abundance in the early phase of
metal enrichment in the Universe.
In particular, extremely metal-poor (EMP) stars with metallicities
$\abH{Fe} < -3$ are considered to form in gas clouds enriched by only 
a few progenitors \citep{Audouze95, Ryan96, Cayrel04}.\footnote{The 
logarithmic number abundance ratio between elements A and B relative to solar one 
${\rm [A/B]} = \log (n_{\rm A}/n_{\rm B}) - \log (n_{\rm A}/n_{\rm B})_{\bigodot}$ 
is often used to measure the metal content and peculiarity of
elemental abundance ratio of a star.
We also use the logarithmic abundance 
$\abA{M} = \log \epsilon ({\rm M}) = 12 + \log (n_{\rm M}/n_{\rm H})$
of an element M.
We hereafter use the solar abundance of \citet{Asplund09}.}
Inversely, we can indirectly see the nucleosynthesis of the progenitors
from the metallicities and elemental abundances of EMP stars.
This approach to chemical evolution in the early Universe 
is called {\it galactic archaeology} \citep{Salvadori07, deBennassuti14, deBennassuti17, Graziani15, Hartwig18, Komiya20}.
Also, for the ancient stars to be observed in the present day,
they should be low-mass ($<0.8 \ \Msun$).
The metallicity distribution of EMP stars gives a constraint on their initial mass functions (IMFs)
with different metallicities.

Another important aim of this study is to search for the first generation of 
metal-free (Population III or Pop III) stars that first modify  
cosmic structure formation through their radiative and SN feedback.
So far, metal-free stars have not been observed although a large number 
($\sim 5000$) of metal-poor stars have been identified in the Milky Way halo and Local 
Group dwarf galaxies in large survey campaigns and follow-up spectroscopic observations.
This indicates that Pop III stars are predominantly massive.
Numerical studies \citep{Bromm99, Abel02, Yoshida03} also predict that massive Pop III stars with
$\MPopIII \sim 10$--$1000 \ \Msun$ form in the metal-free gas clouds
hosted by low-mass ($\sim 10^6 \ \Msun$) dark matter (DM) halos (minihalos; MHs) at 
redshift $z\gtrsim 6$.
The fragmentation of the pristine clouds is significantly reduced due to the lack
of efficient gas coolants.

On the other hand, the survival of stars with non-zero metallicities indicates
that the additional gas cooling due to heavy elements induces the fragmentation of their 
parent clouds.
In particular, thermal emission cooling of dust grains becomes dominant at high
densities $\nH \sim 10^{12}$--$10^{14} \ \percc$, corresponding to small
Jeans masses $\Mjeans \sim 0.01$--$0.1 \ \Msun$, which may be indicative of low-mass star formation
\citep{Omukai00, Omukai05, Schneider02, Schneider03, Schneider06, Schneider12}.
Several authors have studied dust-induced fragmentation,
but they assume that the metal-poor clouds have the
solar elemental abundance ratio and the same composition and size distribution of grains as
in the local interstellar medium (ISM) \citep{Omukai00, Ritter15, Smith15, SafranekShrader16}
\citep[see however][]{Schneider06}.

Following the discovery of the EMP star \Caf \ by \citet{Caffau11}, 
with a total metallicity of only $Z = 4.5\E{-5} \ \Zsun$, 
\citet{Schneider12Caf} showed that its surface elemental 
abundances suggest that its birth cloud was enriched by the metal 
and dust yields of normal core-collapse SNe (CCSNe) and that the gas cooled 
through silicate dust cooling and fragmented. 
The same method was then applied by \citet{Marassi15} to investigate the 
origin of \Kel, a carbon-enhanced star with an upper limit on its surface 
iron abundance of only $\abH{Fe} < -7.1$ \citep{Keller14}. 
It was suggested that, similarly to other carbon enhanced metal-poor (CEMP) stars with $\abH{Fe}\lesssim -3$
\citep[so-called Group II and III of CEMP stars;][]{Yoon16},\footnote{Instead, Group I stars are 
distributed mostly at intermediate metallicities, with $\abH{Fe} \sim -2$, and 
show also s-process element enhancement. 
Because of this, their carbon-enhancement is believed to originate from mass transfer 
from an AGB companion in a binary system \citep{Suda04}.
Alternatively, Group I stars could form in regions that have been previously 
enriched by the explosion of rotating massive stars, 
that could produce C/N and s-process elements in their envelopes 
\citep{Meynet06, Choplin19}.}
this star could originate from dust-cooling and fragmentation of 
a collapsing gas cloud previously enriched by the yields of faint SNe. 
In faint SNe, mixing and fall back of the innermost layers of the ejecta 
into the central compact remnant can yield the C-enhanced abundance pattern of 
ejected materials. 
Since $^{56}$Ni, main source of $\gamma$-ray photons through its radioactive decay, 
is also depleted, this type of SNe is called faint SNe \citep{Umeda03}. 
In addition, \citet{Marassi14, Marassi15} show that the carbonaceous grains 
are produced from faint SNe and could be responsible for the formation of \Kel. 
Hence, it was suggested that the distinctive elemental abundances between C-normal and 
C-enhanced EMP stars point to a different type of SN explosion as the 
main formation sites of their surface elemental abundance, but to a common formation 
pathway based on either silicates or carbon dust cooling \citep{Marassi15}. 
This scenario was further investigated by \citet{Chiaki17}, who showed that the observed 
lower-limits of carbon abundance $\abAcr{C} \sim 6$ for CEMP \citep[for reference, 
the solar abundance is $\abAsun{C} = 8.43$;][]{Asplund09} and of iron abundance of $\abH{Fe}_{\rm cr} \sim -5$ 
for C-normal EMP (CN-EMP) stars confirm that gas cooling by dust driven cooling and fragmentation
in their birth clouds is required, and that the dominant grain species are carbon and silicates, respectively.

Although these models successfully reproduce the {\it relative} element-to-element abundances  
of observed stars, the studies for the {\it absolute} abundance of metals/grains in enriched
clouds have so far been limited to normal CCSNe.
In our previous study \citep[][hereafter \citetalias{Chiaki19}]{Chiaki19}, 
we followed the metal enrichment from a normal CCSN with $\abFe{C} = 0.18$.
We found that only a fraction $\ffb \sim 0.4$\%
of metals return to the MH because
the SN shell interacts with dense cosmological filaments and loses its
energy through radiative cooling.
Still, the metallicity of an enriched cloud $\abH{Fe} = -3.62$
is consistent with the metallicity range of observed C-normal stars.\footnote{In \citetalias{Chiaki19}, 
we presented the iron abundance of the recollapsing cloud as $\abH{Fe} = -3.42$.
In the simulation, we used the solar metallicity $Z_{\bigodot} = 0.01295$ as in this work,
and the metallicity of the recollapsing cloud was $Z = 3.4\E{-6} = 2.6\E{-4} \ \Zsun$.
However, when we converted $Z$ from 
$\abH{Fe} = 12 + \log (\abX{Fe} Z / \mu _{\rm Fe} \XH) - \abAsun{Fe}$, 
we used $\Zsun = 0.02$.
If we use $\Zsun = 0.01295$, $\abH{Fe}$ is consistently estimated to be $-3.62$.}
Mainly silicate grains ejected from the SN induce gas cooling in the enriched cloud
and fragmentation occurs.
Hence, the scenario proposed in \citetalias{Chiaki19} could explain the origin of C-normal stars
from a single CCSN.

For faint SNe, the ejected metal mass is smaller than for normal CCSNe because  
the ejecta partially falls back.
Also, faint SNe exploding in a spherically symmetric manner have smaller
explosion energies than normal CCSNe with the same progenitor mass.
This can affect the return fraction of metals $\ffb$.
Therefore, it is necessary to investigate whether the enriched clouds have C abundances consistent 
with the observed level ($\abA{C} \gtrsim 6$).

In the present study, we follow the metal enrichment from a faint SN 
and the gravitational collapse of an enriched cloud
with a set of cosmological simulations.
To quantify the absolute metal and dust abundances in the enriched cloud, 
for the first time we employ a nucleosynthesis/nucleation model of Pop III faint SNe 
produced by \citet[][hereafter \citetalias{Marassi14}]{Marassi14}.
The metal mass and explosion energy are consistently derived so that 
the model reproduces the elemental abundance of the CEMP star \Kel \ \citep{Keller14} 
for given progenitor masses.
We then follow the gravitational collapse of the enriched cloud with all relevant cooling/heating processes
including carbon grain cooling and chemical reactions as well as growth of dust grains \citep{Chiaki15}
to determine whether the gas cloud is able to fragment, allowing the formation of low-mass and long-lived stars,
such as \Kel .

The structure of this paper is as follows:
In Section \ref{sec:method}, we describe our numerical methods.
Then, the results are presented in Section \ref{sec:results}.
We discuss the observability of stars forming in the simulated clouds
enriched by faint SNe and other issues in Section \ref{sec:discussion}.
Finally, the paper is concluded in Section \ref{sec:conclusion}.
Throughout the simulations, we adopt the cosmological parameters $\Omega _{\rm m} = 0.3089$,
$\Omega _{\rm CDM} = 0.2603$, $\Omega _{\Lambda} = 0.6911$, and $H_0 = 67.74 \ {\rm km \ s^{-1} \ Mpc^{-1}}$
\citep{Planck2015}.
We run the simulations in comoving coordinates but we describe physical quantities in proper coordinates
throughout this paper, unless otherwise specified.
All the figures in this paper are created with the {\sc yt} toolkit \citep{yt}.\footnote{
\url{https://yt-project.org/}.
The version and the script used in this work are respectively available at \\
\url{https://github.com/genchiaki/yt/tree/metal-dust}, \\
\url{https://github.com/genchiaki/Analysis_CEMP}. \\
The simulation data will be shared on reasonable request to the authors.}

\section{Numerical models}
\label{sec:method}

Since the numerical method is nearly the same as \citetalias{Chiaki19}, 
we briefly describe it in Sections \ref{sec:basic_setup} and \ref{sec:simulations_of_cemp_star_formation}. 
We detail a faint SN model that is, for the first time, included in
three-dimensional simulations of this work in Section \ref{sec:sn_models}.

\subsection{Basic setup}
\label{sec:basic_setup}

We perform cosmological simulations with the adaptive mesh refinement (AMR)
hydrodynamics code {\sc enzo} \citep{Bryan14}.\footnote{\url{http://enzo-project.org/}. \\
The version used in this work is available at \\ \url{https://github.com/genchiaki/enzo-dev/tree/metal-dust}.}
The dynamics of DM is followed with an $N$-body particle-mesh solver
\citep{Efstathiou85, Bryan97}.
The hydrodynamics equations are solved with the piecewise parabolic method (PPM)
in an Eulerian frame \citep{Woodward84, Bryan95}
and a Harten-Lax-van Leer-Contact (HLLC) Riemann solver, which accurately captures hydrodynamical
shocks and computes advection of chemical species across contact discontinuities.
We set the controlling parameters for the flux calculation and 
interpolation of field values between computational grids as
${\tt FluxCorrection} = 2$ and ${\tt ConservativeInterpolation} = 0$ to minimize 
numerical errors.
Although errors for the total metal mass accumulates up to $\sim 7$\% at $\sim 500$ Myr
after a SN explosion even with the above setting, 
this is below the typical observational error of stellar abundances ($\sim 0.1$ dex).

Computational cells are progressively refined by a factor of two in space
when satisfying the following criteria:
\begin{itemize}
\item[(i)] The baryon mass in a cell exceeds $3 m_{\rm b,0} \times 2^{-0.2 l}$ on a refinement level $l$,
where $m_{\rm b,0}$ is the mean baryon mass on the root grid.
\item[(ii)] The DM particle mass contained by a cell exceeds $3 m_{\rm dm,0}$,
where $m_{\rm dm,0}$ is the mean DM mass on the root grid.
\item[(iii)] The local Jeans length $\ljeans$ is resolved less than 64 cells.
\end{itemize}
The negative coefficient $-0.2$ in the spectral index of the criterion means (i) the super-Lagrangian 
refinement criterion for the gas component while the criterion (ii) the Lagrangian for the DM.
When the baryon density starts to increase in the run-away collapse phase,
cells are refined mostly on the criterion (iii).
This criterion warranties that the local Jeans length is resolved sufficiently 
to prevent spurious fragmentation \citep{Truelove97, Turk12}.

We generate the initial conditions of the simulations in a periodic box with a side of 300 kpc (comoving)
with {\sc music} \citep{Hahn11}.
We initially run a DM-only simulation with a base resolution $64^3$ and identify
the most massive halo with a mass $3.8\E{6} \ \Msun$ at redshift $z = 10$
with a friends-of-friends (FOF) algorithm.
By initially refining the halo region with two additional AMR levels, i.e., with higher spatial resolution by 
a factor of four, we restart the simulation adding the baryon component.
With this zoom-in strategy, the effective resolution is $256^3$ and the minimum DM particle
mass is $53.4 \ \Msun$.

\begin{figure*}
\includegraphics[width=\textwidth]{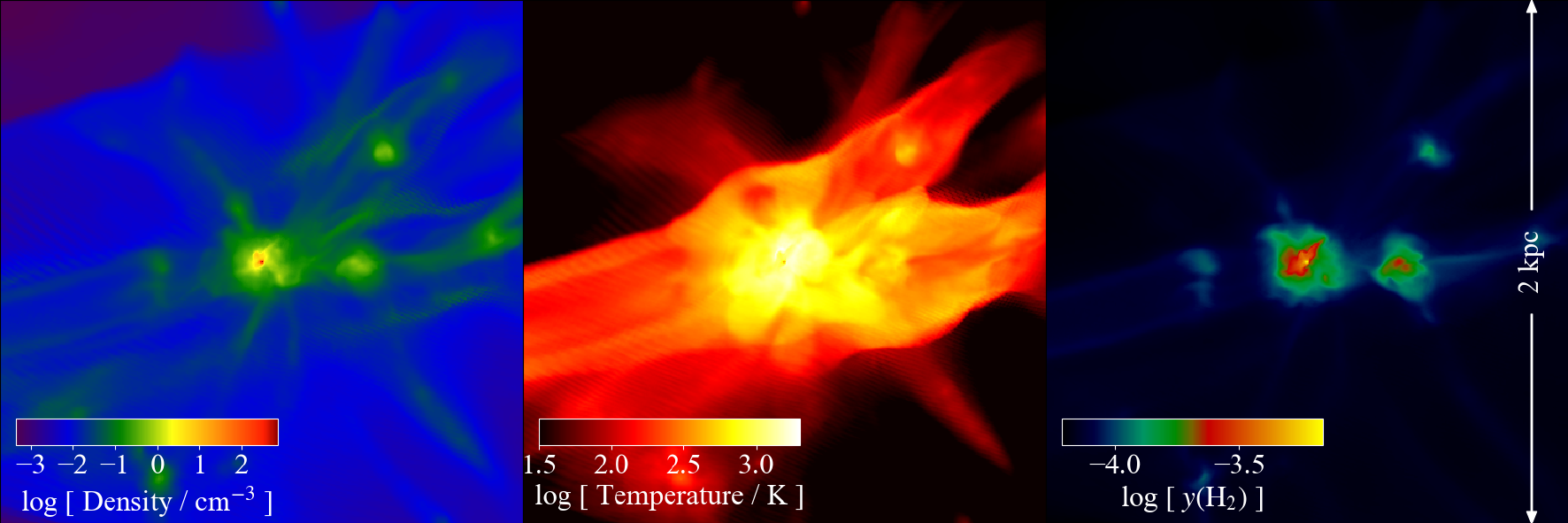}
\caption{
Density-weighted projection of density, temperature, and H$_2$ abundance of the minihalo just before Pop III star formation
(redshift $z=12.1$) in a box with a side 1 kpc centered on the density maximum.
The horizontal and vertical axes are respectively parallel to the major and minor axes of the 
momentum of inertia of the region with densities above $0.1 \ \percc$.
}
\label{fig:snapshots_ini}
\end{figure*}

\subsection{Simulations of CEMP star formation}
\label{sec:simulations_of_cemp_star_formation}

In this section, we describe the numerical methods that follow the formation
of a Pop III star (Section \ref{sec:pop_iii_star_formation}), its radiative and SN feedback (Section \ref{sec:radiation_and_supernova_feedback_from_a_pop_iii_star}), and 
recollapse of an enriched cloud in the MH (Section \ref{sec:collapse_of_the_enriched_clouds}).

\subsubsection{Pop III star formation}
\label{sec:pop_iii_star_formation}

The main coolant of a primordial cloud is hydrogen molecules (H$_2$) and,
in some cases, hydrogen deuteride molecules (HD).
To calculate the fractions and cooling rates of H$_2$ and HD, we solve the non-equilibrium chemistry 
with a modified version of a chemistry/cooling library {\sc grackle} 
\citep[][\citetalias{Chiaki19}]{Smith17}.\footnote{\url{https://grackle.readthedocs.io/}. \\
The version used in this work is available at \\ \url{https://github.com/genchiaki/grackle/tree/metal-dust}.}
We solve a chemical network of 49 reactions for 15 primordial species, 
e$^-$, H$^+$, H, H$^-$, H$_2^+$, H$_2$,
D$^+$, D, D$^-$, HD$^+$, HD, 
He, He$^+$, He$^{2+}$, and HeH$^+$.
This chemical network includes 
the collisional ionization/recombination of H/He and
formation/dissociation of H$_2$/HD molecules.
We compute the rates of radiative cooling:
inverse Compton, bremsstrahlung, H/He line transition, 
H$_2$ ro-vibrational transition, and HD rotational transition cooling.
When H$_2$ molecules form, the binding energy (4.48 eV per molecule) is converted to the thermal
energy \citep[see][]{Omukai00}.
The continuum opacity $\kappa _{\rm p}$ of the primordial component is taken from \citet{Mayer05}.
Throughout this paper, the mass fraction of hydrogen nuclei is $\XH = 0.76$ and number fraction of
deuterium relative to hydrogen nuclei is $y_{\rm D} = 3.4\E{-5}$.

We then simulate the formation of a Pop III star.
In reality, a star forms from a hydrostatic core with a central density $\nH \sim 10^{19} \ \percc$ where 
gas cooling from the endothermic reaction of hydrogen molecular dissociation becomes ineffective
\citep{Larson69, Penston69}.
In this work, to save the computational cost,
we put a Pop III star particle, representing a single star, with the criteria:
\begin{itemize}
\item[(i)] gas density exceeds $10^6 \ \percc$,
\item[(ii)] gas flow is convergent ($\nabla \cdot {\vb v} < 0$),
\item[(iii)] the cooling time is less than the dynamical time,
\item[(iv)] H$_2$ fraction exceeds a critical value ($10^{-3}$).
\end{itemize}
Fig. \ref{fig:snapshots_ini} shows the snapshot of the MH just before the 
time $t_{\rm form}$ of Pop III star formation.
At this time ($z=12.1$), the DM and baryon mass is 
$\Mhalodm = 1.53\E{6} \ \Msun$ and $\Mhalob = 2.57\E{5} \ \Msun$, respectively,
within the virial radius $\Rhalo = 288$ pc.
The H$_2$ fraction reaches $\sim 10^{-3}$ through the H$^-$ process \citep{Omukai00}
and the temperature at the center decreases because of H$_2$ cooling.
In this region, the above criteria for Pop III star formation are satisfied.

From the time $t_{\rm form}$, we run three simulations, in
each of which we put a single Pop III star with masses $13.5$, $50$, and $80 \ \Msun$
in the MH
(hereafter called the run {\tt F13}, {\tt F50}, and {\tt F80}, respectively).

\subsubsection{Radiation and supernova feedback from a Pop III star}
\label{sec:radiation_and_supernova_feedback_from_a_pop_iii_star}

During the main sequence of the Pop III star,
we solve the radiative transfer with a module {\sc moray} \citep{Wise11}. 
The ionization photon emission rate $\QH$ and the lifetime of the stars $\tlife$ 
are taken from \citet{Schaerer02} and reported in Table \ref{tab:PopIII}.
We calculate the H, He, and He$^+$ ionization rates by integrating the number of 
photons consumed by the atoms from the location of the star particle to each computational cell 
with the cross-section of \citet{Verner96}.
The H$_2$ dissociation rate is computed from the column density of H$_2$
with a self-shielding function of \citet{Draine96}.
With these rates, the photon reactions
are consistently calculated, coupled with the non-radiative reactions.

After the time $\tlife$, we uniformly add 
explosion energy and metals within a sphere with a radius $\Rs = 10$ pc
centered on the star particle, mimicking the end of the free-expansion phase.
We assume that the explosion energy has been completely converted to the thermal energy.
We also assume that the ejecta has been completely mixed during the free-expansion phase
and its elemental composition is uniform.
The explosion energy $\Esn$ and mass of ejected metals/grains of our models are described 
in Section \ref{sec:sn_models} and listed in Table \ref{tab:SN}.

\begin{table}
\begin{minipage}{\columnwidth}
\caption{Pop III models}
\label{tab:PopIII}
\begin{tabular}{cccc}
\hline
$^{1}$Run       & $^{2}$$\MPopIII$ & $^{3}$$\tlife$ & $^{4}$$\QH$                \\
                &       $[\Msun]$  &       [Myr]    & [$10^{48} \ {\rm s^{-1}}$] \\
\hline \hline                                                             
{\tt F13}       &       $13.5$     &      $11.8$    & $1.09$ \\
{\tt F50}       &       $50$       &      $3.49$    & $34.6$ \\
{\tt F80}       &       $80$       &      $3.03$    & $84.9$ \\
\hline
\end{tabular}
\medskip \\
Note --- (1) ID of runs.
(2--4) mass, lifetime, and ionizing photon emission rate of our Pop III star models.
\end{minipage}
\end{table}

\begin{table*}
\begin{minipage}{\textwidth}
\caption{SN models}
\label{tab:SN}
\begin{tabular}{cccccccccccc}
\hline
$^{1}$Run        & $^{2}$$\Esn$          & $^{3}$$M_{\rm cut}$ & $^{4}$$\Mmet$   & $^{5}$$\abM{C}$ & $^{6}$$\abM{O}$ & $^{7}$$\abM{Fe}$   & $^{8}$[C/Fe]  & $^{9}$$M_{\rm AC}$   & $^{10}$${\cal D}_{\rm AC}$ & $^{11}$$r_{\rm AC, cool}$ \\
                 &       [$10^{51}$ erg] &        $[\Msun]$    &       $[\Msun]$ &       $[\Msun]$ &       $[\Msun]$ &       $[\Msun]$    &               &       $[\Msun]$      &                            &        $[\mu m]$          \\
\hline \hline                                                                                                                                                            
{\tt F13}        & $0.5$                 & $1.7$               & $0.119$         & $0.079$         & $0.039$         & $1.06\E{-6}$       & $4.62$        & $1.70\E{-2}$         & $1.49\E{-8}$               & $0.096$            \\
{\tt F50}        & $2.6$                 & $11 $               & $3.54$          & $0.989$         & $2.551$         & $1.47\E{-5}$       & $4.57$        & $3.89\E{-4}$         & $4.11\E{-10}$              & $0.012$            \\
{\tt F80}        & $5.2$                 & $22.5$              & $4.31$          & $1.089$         & $3.213$         & $1.05\E{-5}$       & $4.75$        & $3.74\E{-2}$         & $2.51\E{-8}$               & $0.064$            \\
\hline
\end{tabular}
\medskip \\
Note --- (1) ID of runs.
(2) explosion energy.
(3) mass cut.
(4--7) metal, carbon, oxygen, and iron mass in ejecta.
(8) carbon-to-iron abundance ratio.
(9) mass of amorphous carbon (AC).
(10) mass ratio of AC to gas.
(11) characteristic radius of AC grains (see text).
\end{minipage}
\end{table*}

\subsubsection{Collapse of the enriched clouds}
\label{sec:collapse_of_the_enriched_clouds}

To include the chemical/thermal evolution of clouds enriched by faint SNe, 
we solve 40 reactions of 19 metal species:
C$^+$, C, CH, CH$_2$, CO$^+$, CO, CO$_2$, O$^+$, O, OH$^+$, OH, H$_2$O$^+$, H$_2$O,
H$_3$O$^+$, O$_2^+$ , O$_2$, Si, SiO, and SiO$_2$.
In our cooling model, we include C {\sc ii}, C {\sc i}, and O {\sc i} fine-structure 
transition line cooling and CO, OH, and H$_2$O molecular rotational transition line cooling.
The optical depth in each transition line is calculated with the Sobolev length
approximation.
To reduce the computational cost, we estimate the local velocity gradient to be 
$|\nabla \cdot {\vb v}| = 1/3 \tff$, where $\tff = \left(3 \pi / 32 G \rho \right)^{1/2}$ 
is the free-fall time for a density $\rho$ \citep{Omukai00}.

We model the chemistry and cooling of dust grains most carefully.
In this work, we consider 7 grain species: 
alumina (Al$_2$O$_3$),
metallic iron (Fe), 
magnetite (Fe$_3$O$_4$), 
enstatite (MgSiO$_3$), 
forsterite (Mg$_2$SiO$_4$), 
amorphous carbon (AC), and
silica (SiO$_2$)
with arbitrary grain size distributions.
Further, we include grain growth (accretion of gas-phase metals onto grains)
as chemical reactions with a reaction rate given by 
the geometrical cross-section of the grains multiplied by the thermal velocity of 
metal molecules \citep{Kozasa87}.
The cooling rates, H$_2$ formation rates on grain surfaces, continuum opacity, and
grain growth rates are calculated for each species.
We calculate the continuum optical depth as
$\tau _{\rm cont} = (\kappa _{\rm p} \rho + \sum _i \kappa _i \rho _i) \ljeans$
with an opacity $\kappa _i$ \citep{Nozawa08} and density $\rho _i$ of a grain species $i$.

We terminate the simulations at a density $\nH \sim 10^{16} \ \percc$ above
which further cloud fragmentation does not occur.
In this optically thick region, radiative cooling becomes ineffective, and thus
stable hydrostatic cores (first cores) form \citep{Larson69}.

We run the simulations on the PACE cluster in Georgia Institute of Technology on
3 nodes, using 28 cores/node.
The maximum AMR level eventually reaches 32--34, corresponding to a spatial size of $\sim 0.01$ au.
Even protostars with a size $\sim 1$ au can be resolved.
The runs {\tt F13}, {\tt F50}, and {\tt F80} take 69, 47, and 41 days with total computational
costs of 140,000, 94,000, and 83,000 core-hours, respectively.

\begin{figure}
\includegraphics[width=\columnwidth]{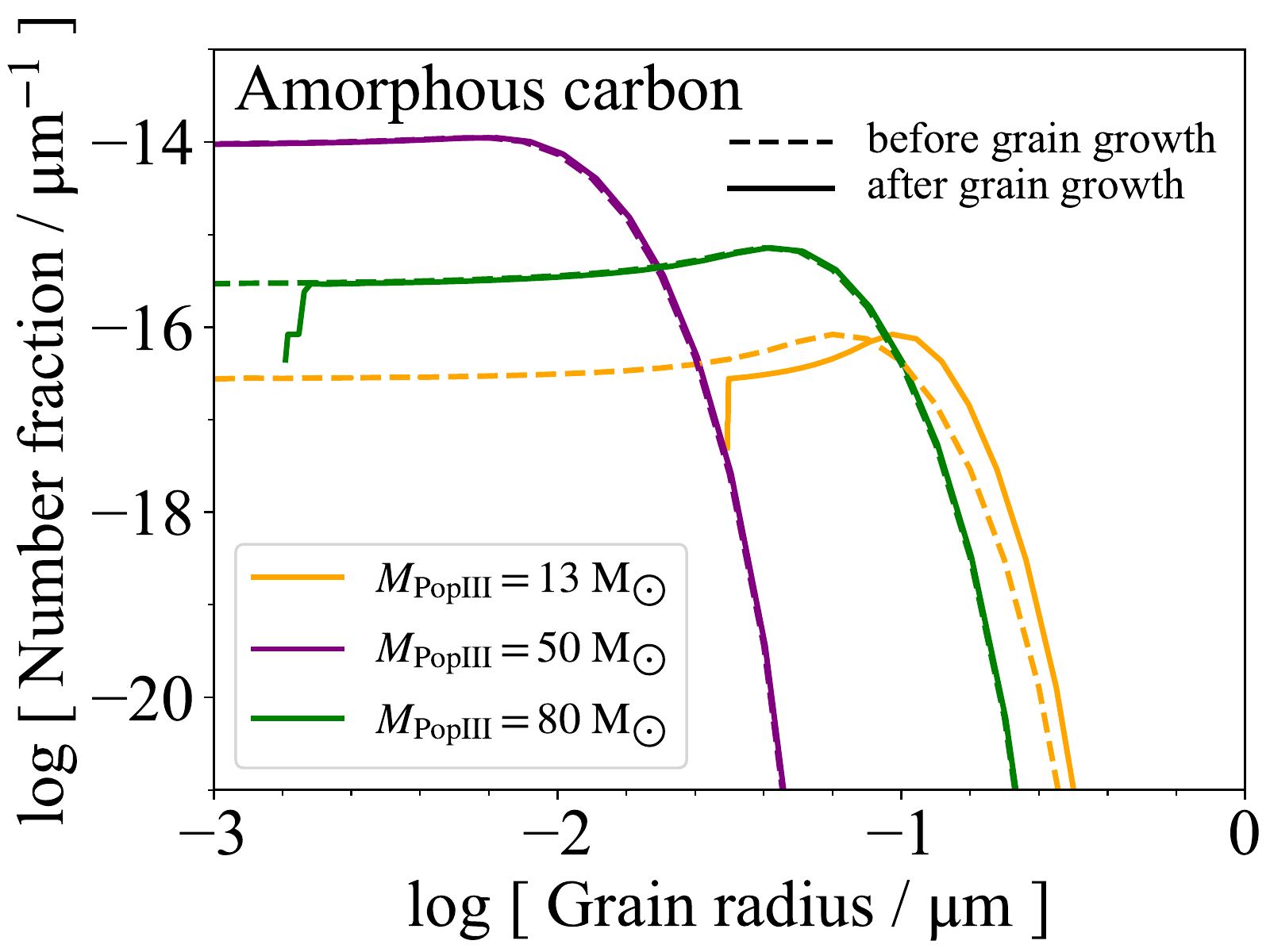}
\caption{
Distribution of the number fraction of amorphous carbon (AC) grains relative to hydrogen nuclei
against grain radii in the clouds enriched by Pop III SNe
with progenitor masses $13 \ \Msun$ (orange), $50 \ \Msun$ (purple) and $80 \ \Msun$ (green).
The dashed and solid curves denote the size distribution of grains in envelope and center
of the clouds before and after grain growth occurs, respectively.
}
\label{fig:dist}
\end{figure}

\subsection{SN models}
\label{sec:sn_models}

We adopt faint SN models calculated by \citetalias{Marassi14}.
In these models, explosion energy $\Esn$, $^{56}$Ni mass, and initial
ejection velocity are chosen so that the abundance ratios 
of heavy elements are fitted with the elemental abundance of the most
iron-poor CEMP star, \Kel \ \citep{Keller14}, for each progenitor mass.
The mass cut, $\Mcut$, depends on the initial ejection velocity.
The total mass of synthesized metals, $\Mmet$, is $0.119$--$4.31 \ \Msun$.
Because of the fall back of the inner layers of the ejecta, 
the metal mass $0.119 \ \Msun$ for {\tt F13} is smaller than that of
normal CCSNe with the same progenitor mass ($0.3$--$1 \ \Msun$).
The masses of C, O, and Fe range between $0.079$--$1.089 \ \Msun$, $0.039$--$3.213 \ \Msun$, 
and $1.06\E{-6}$--$1.05\E{-5} \ \Msun$, respectively.
This gives a C-enhanced abundance ratio of $\abFe{C} = 4.57$--$4.75$.
In these spherical explosion models, the explosion energy for {\tt F13} is smaller ($0.5\E{51}$ erg) 
than the normal CCSN model with the same progenitor mass used in \citetalias{Chiaki19} by a factor of two.

\begin{figure*}
\includegraphics[width=\textwidth]{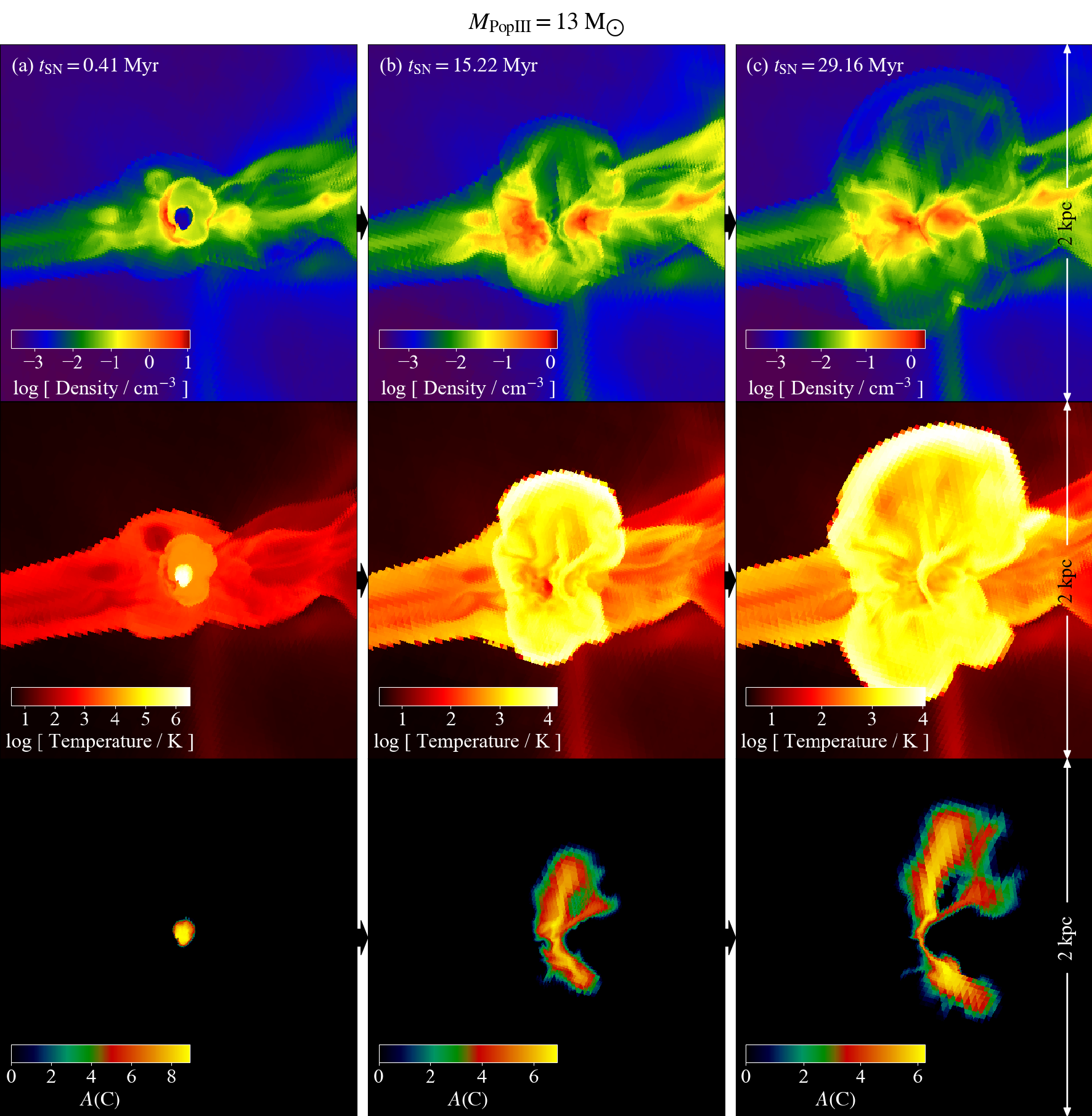}
\caption{
Slice of density, temperature, and C abundance for a progenitor model with a mass $\Mpr = 13 \ \Msun$
at the time $t_{\rm SN} = 0.41$ Myr (column a), 
15.22 Myr (column b), and 29.16 Myr (column c)
after the SN explosion
in a box with a side 2 kpc centered on the centroid of the MH.
The axes of the windows are the same as Fig. \ref{fig:snapshots_ini}.
}
\label{fig:snapshots_exp_M13}
\end{figure*}

\begin{figure*}
\includegraphics[width=\textwidth]{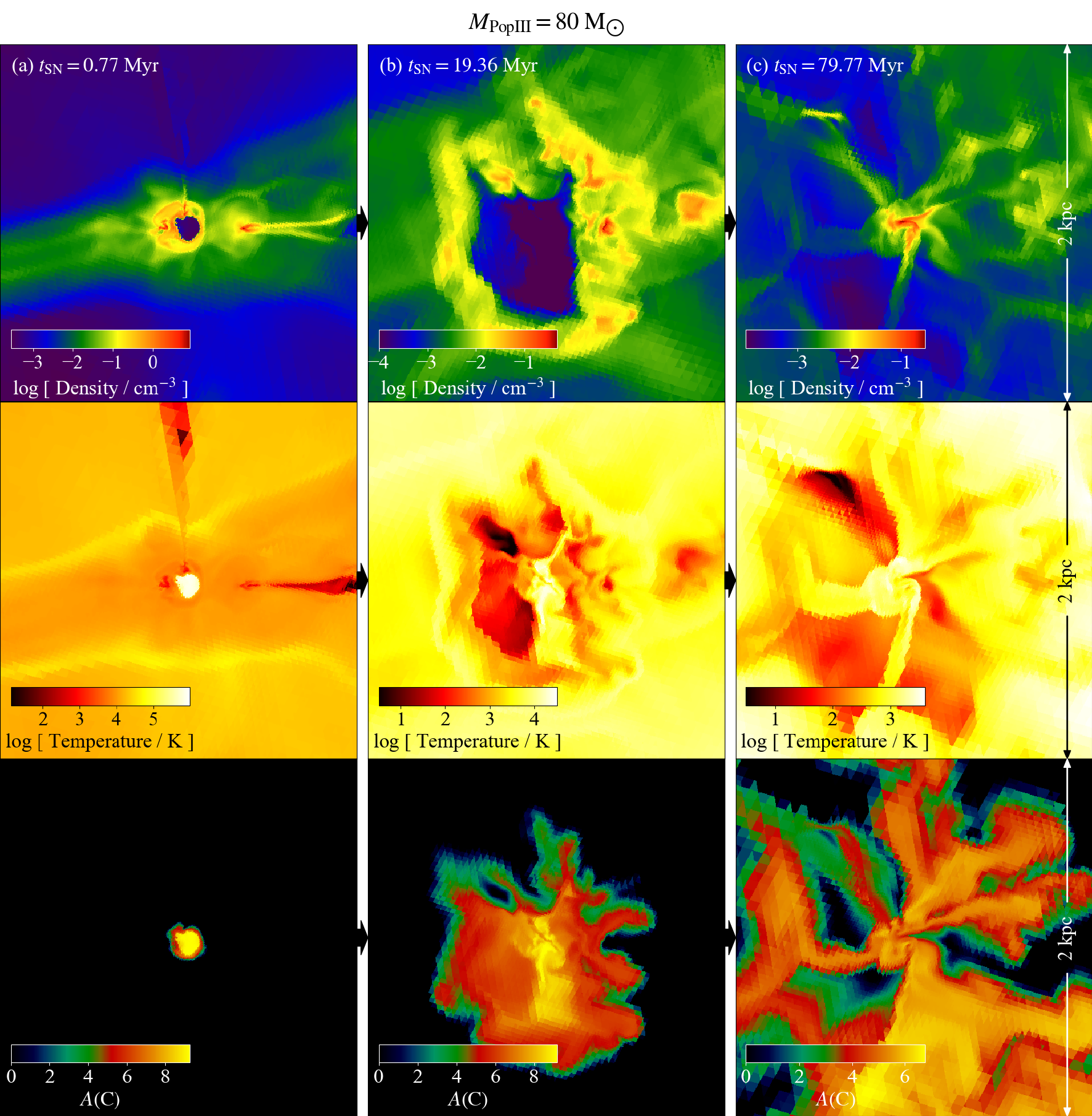}
\caption{
Same as Fig. \ref{fig:snapshots_exp_M13} but for a progenitor mass $80 \ \Msun$ at the time 
$\tsn = 0.77$ Myr (column a), 19.36 Myr (column b), and 79.77 Myr (column c)
from the SN explosion.
}
\label{fig:snapshots_exp_M80}
\end{figure*}

\citetalias{Marassi14} follow the formation/destruction of dust grains in faint SNe.
First, dust grains form in the expanding ejecta at the time $\sim 100$ days after the SN explosion.
At the time when gas temperature decreases below the sublimation temperature of grains ($\sim 2000$ K),
seed clusters form through aggregation of monomers and grow by accreting the gas-phase species.
In C-rich ejecta, AC is the dominant species while the mass fraction of other species is at most $10^{-5}$.
The newly formed grains have an approximately log-normal size distribution function with a mean radius
that depends on the C density and temperature at the grain formation time.
Then, \citetalias{Marassi14} follow the destruction of dust grains in SN remnants.
When the ejecta sweeps up the circumstellar gas with a mass comparable to the ejecta mass,
reverse shocks propagate to the region where grains have formed.
Through the collision of high energy ions, 
a fraction of the atoms on grain surfaces returns to the gas phase.
Dust destruction occurs at $\sim 1000$ years after the SNe.
This should occur in the three-dimensional simulations but
we do not include the dust destruction to save the numerical cost.
Instead, we use the dust model of \citetalias{Marassi14} as an initial condition.

The efficiency of grain destruction depends on the density $\rhoamb$ of ambient gas around a SN.
\citetalias{Marassi14} calculate the grain properties for $\rhoamb = 10^{-25}$--$10^{-23} \ {\rm g} \ \percc$
for each progenitor mass.
We employ the model for $\rhoamb = 10^{-24} \ {\rm g} \ \percc$.
Table \ref{tab:SN} shows the mass $\abM{AC}$ of AC grains produced by
faint SNe after the formation/destruction processes. 
While the fraction of C locked up into AC grains is 4 and 20\% for {\tt F13} and {\tt M80}, respectively,
it is only 0.04\% for {\tt F50}.
The ejecta density is larger for {\tt F50} than for the other progenitor masses, and thus
C atoms are oxidized to form CO molecules more rapidly than condensed into AC grains.
We can expect that the low abundance of AC prevents the fragmentation of a cloud enriched by the SN.
The dashed curves of Fig. \ref{fig:dist} show the size distribution of AC grains after the destruction.
When grains are sputtered, their size decreases.
A tail appears on the smaller end of the initial log-normal size distribution.
Following \citet{Chiaki15, Chiaki17}, we also estimate a characteristic grain radius given by 
$r_{\rm AC, cool} = \langle r^3 \rangle _{\rm AC} / \langle r^2 \rangle _{\rm AC} = 0.012$--$0.096 \ \um$
from the grain properties in our SN models,
where $\langle r^n \rangle _{\rm AC} = \int r^n \varphi _{\rm AC} (r) dr$ 
is the $n$-th moment of size distribution function $\varphi _{\rm AC} (r)$ at a grain radius $r$
normalized to unity.

\begin{figure}
\includegraphics[width=\columnwidth]{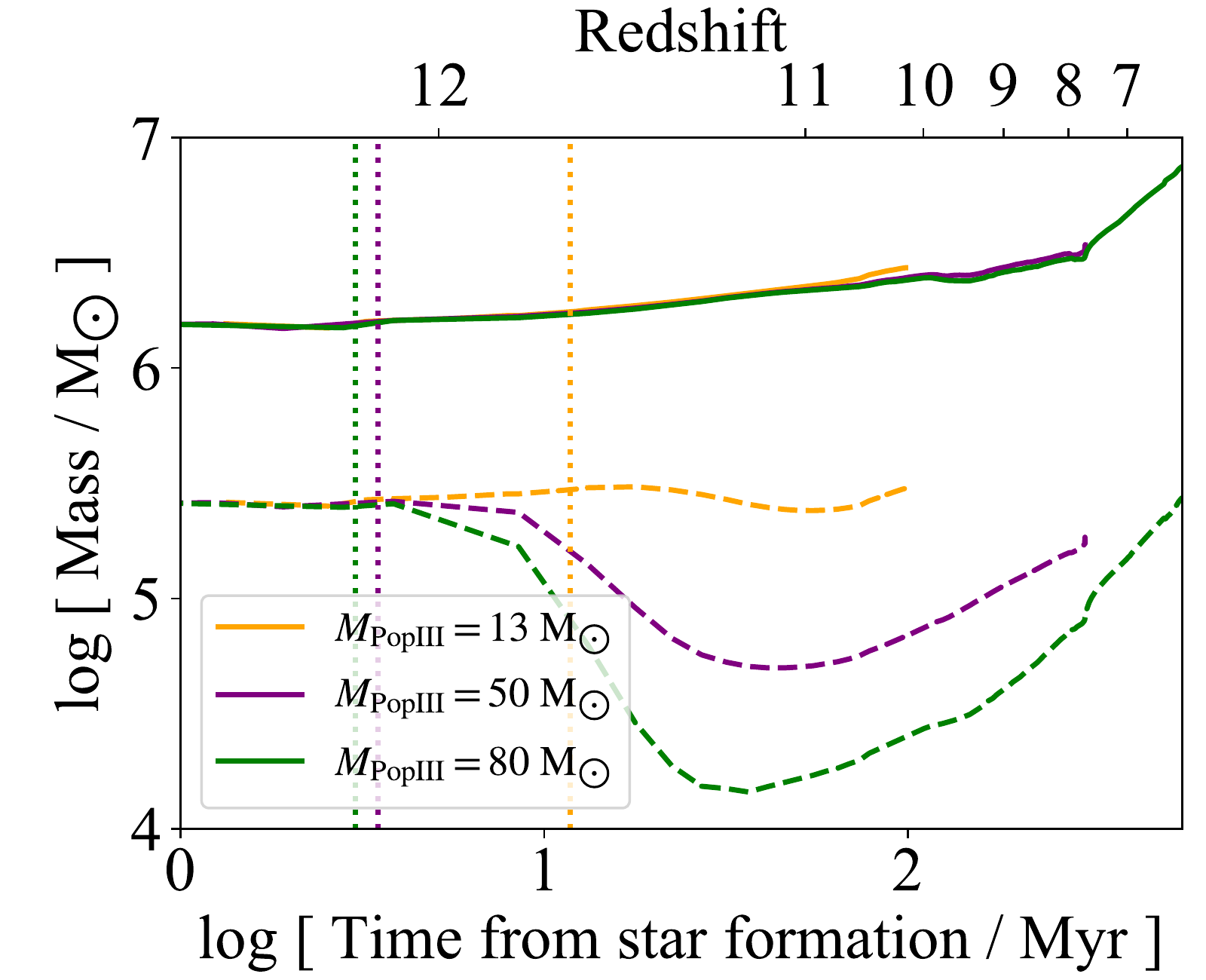}
\caption{
Temporal evolution of DM and baryon mass within the virial radius of the MH 
(solid and dashed curves, respectively) from the time of Pop III star formation.
The vertical dotted lines show the time of SN explosions
for progenitor masses $13 \ \Msun$ (orange), $50 \ \Msun$ (purple) and $80 \ \Msun$ (green).
}
\label{fig:tm}
\end{figure}

\section{Results}
\label{sec:results}

\subsection{Enrichment from faint SNe}

The evolution of SN shells is affected by
the density structure of the H {\sc ii} region created by the ultraviolet (UV) radiation from the progenitor star.
The column (a) of Figs. \ref{fig:snapshots_exp_M13} and \ref{fig:snapshots_exp_M80} shows the density, temperature,
and carbon abundance, $\abA{C}$, just after the SN.
We do not show the plot for {\tt F50} because it is essentially the same as for {\tt F80}.
From these figures, we can see the structure of outer part of H {\sc ii} region which has not
been affected by the SN shock.
For {\tt F13}, the H {\sc ii} region formed around the Pop III star has a radius $\sim 100$ pc
defined by a D-type shock front.
In the region, the density and temperature become 
$\sim 0.01 \ \percc$ and $10^4$ K, respectively, and hydrogen atoms are fully ionized.
For {\tt F50} and {\tt F80}, the ionization front is decoupled from the D-type shock 
at a radius $\sim 100$ pc and reaches $\sim$ kpc beyond the virial radius because of the large ionizing photon 
emission rates $\QH = 3$--$8\E{49} \ {\rm s}^{-1}$.

After the SN explosion, a hot bubble ($\sim 10^7$ K) forms around the Pop III remnant.
Consistent with analytical blastwave models, the temperature declines adiabatically until the timescale for
radiative cooling becomes smaller than the dynamical time of the shell ($\sim 0.1$--$1$ Myr).
Then, the shell loses its energy through radiative cooling and its expansion is driven 
by the hot inner bubble in the pressure-driven snowplough (PDS) phase.
The inertial force from the decelerated ejecta to the outer swept-up
pristine materials drives Rayleigh-Taylor (RT) instabilities and 
metal mixing between the two regions.

After the inner hot bubble loses its thermal energy, the shell expansion is driven
under the conservation of momentum (momentum-conserving snowplough; MCS phase).
Figs. \ref{fig:snapshots_exp_M13}b and \ref{fig:snapshots_exp_M80}b show that the temperature
of the inner region begins to decline at a time $\tsn = 10$--$20$ Myr after the SN explosion.
The RT fingers develop toward less dense regions in the SN shell.
An anti-correlation between density and metal abundance can be seen.
Then, the shell collides with dense clumps at distances of $\sim 100$ pc
on the filaments and loses its kinetic energy.

Finally, the clumps return to the MH as shown in
Figs. \ref{fig:snapshots_exp_M13}c and \ref{fig:snapshots_exp_M80}c.
The collision of the clumps induces the mixing of the metals which remain in 
the halo center.
For all models, the internal enrichment of the MH occurs and
the enriched cloud begins to collapse gravitationally.
We estimate the time $\tret$ when the density of the enriched cloud increases up to $10^{6} \ \percc$ 
to be $86.6$, $304$ and $564$ Myr for {\tt F13}, {\tt F50}, and {\tt F80}, respectively 
(see Table \ref{tab:metallicity}).
For {\tt F50} and {\tt F80} where the H {\sc ii} region expands beyond the virial radius of the MH,
it is expected that the SN shell entirely expands into the void region \citep{Kitayama05, Chiaki18}.
We find that the gas infall continues along the cosmic filaments 
and a part of the shell eventually returns to the MH although it takes longer time than for {\tt F13}.
For {\tt F80}, the time $\tret = 564$ Myr corresponds to redshift $\zret = 6.01$.
Because we do not consider a UV background, our results are only applicable to volumes
that are not ionized by external sources.
However, the same sequence can be generalized and is just as likely to occur slightly earlier
(i.e., $z \sim 8$) in a different large-scale environment.

Fig. \ref{fig:tm} shows the temporal evolution of mass within the virial radius 
from $t_{\rm form}$ to $\tret$.
While the DM mass (solid curves) monotonically increases up to $2.72$--$7.47\E{6} \ \Msun$
through the accretion of DM and a halo merger (at $z \sim 8$),
the baryon mass (dashed curves) decreases from the time of SN explosions (represented by
vertical dotted lines) because the gas is expelled from the MH.
At $\sim 30$ Myr after the explosion, the baryon mass turns to increase 
up to $3.00$--$2.73\E{5} \ \Msun$ 
because the pristine gas continues to be accreted into the MH.
The accretion of pristine gas leads to the dilution of metals. 

\begin{figure*}
\includegraphics[width=\textwidth]{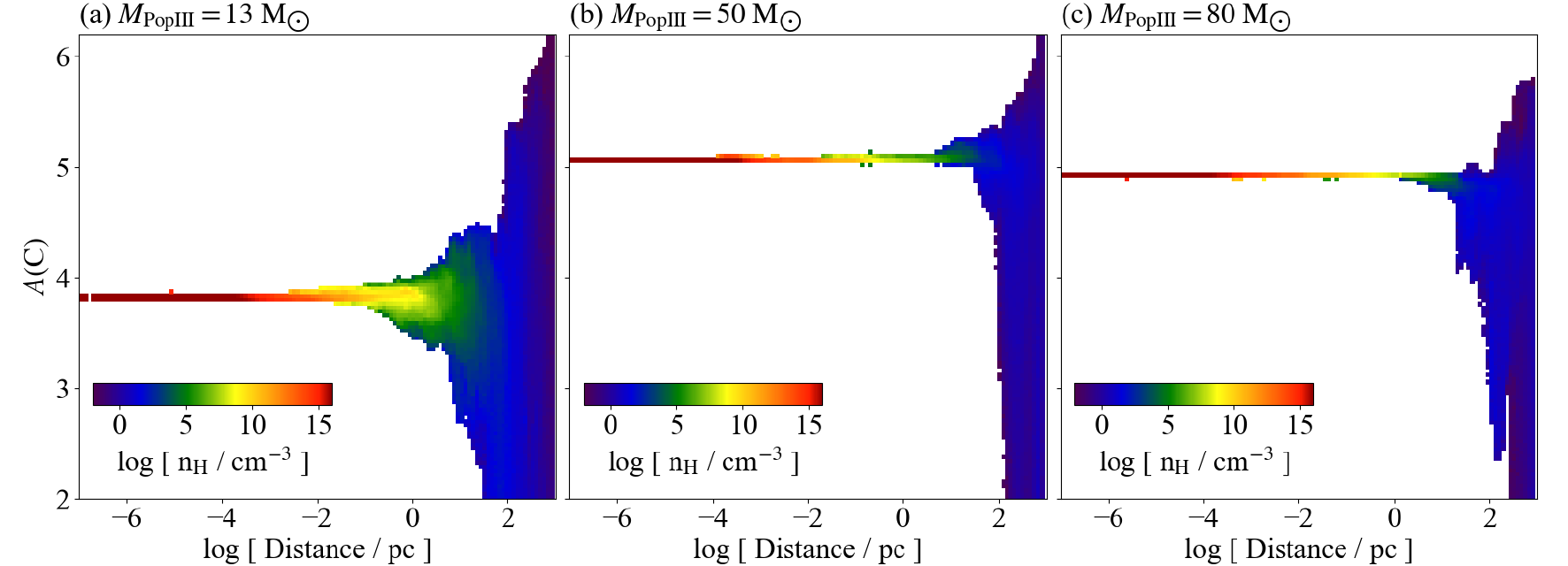}
\caption{
Distribution of C abundance $\abA{C}$ as a function of distance from the density
maximum of the enriched clouds
for progenitor masses (a) $13 \ \Msun$, (b) $50 \ \Msun$ and (c) $80 \ \Msun$
at the time when we terminate the simulations.
}
\label{fig:rZ}
\end{figure*}

Fig. \ref{fig:rZ} shows the distribution of $\abA{C}$ as a function of distance from
the center of the enriched cloud.
In the outskirt of the enriched cloud, the metal fraction as well as C abundance
shows a large scatter.
While the anti-correlation between density and metallicity is still present
in some regions, these two quantities are correlated in the other regions where 
the metal-rich SN shell infalls into the
MH (Fig. \ref{fig:snapshots_exp_M80}c).
The metallicity converges within a radius $\sim 0.1$ pc (for {\tt F13}) 
and $\sim 10$ pc (for {\tt F50} and {\tt F80}) from the cloud center,
indicating the minimum length scale of the fluctuation of metallicity.  
C and Fe abundances in the uniform metallicity region are
($\abH{Fe}$, $\abA{C}$) = ($-9.25$, $3.80$), ($-7.94$, $5.06$), and ($-8.28$, $4.90$)
for {\tt F13}, {\tt F50}, and {\tt F80}, respectively.
The dust-to-gas mass ratio is ${\cal D}_{\rm AC} = 1.49\E{-8}$, $4.11\E{-10}$, and $2.51\E{-8}$
(Table \ref{tab:SN}).
The comparison between the dust cooling rate and the gas compressional heating rate
suggests that clouds with dust-to-gas mass ratios above $\Dcrit = 2.6$--$6.3\E{-9}$ are likely to fragment
into low-mass clumps \citep{Schneider12}.
We now describe the succeeding run-away collapse phase of the enriched clouds, focusing on their
thermal evolution and fragmentation properties.

\begin{table*}
\begin{minipage}{\textwidth}
\caption{Abundance in the enriched clouds and fragmentation properties}
\label{tab:metallicity}
\begin{tabular}{cccccccccccc}
\hline
$^{1}$Run       & $^{2}$$\tret$ & $^{3}$$\zret$ & $^{4}$$\Mhalodm$ & $^{5}$$\Mhalob$  & $^{6}$$\abH{Fe}$ & $^{7}$$\abA{C}$ & $^{8}$${\cal D}_{\rm AC}$ & $^{9}$$r_{\rm AC, cool}$ & $^{10}$$N_{\rm frag}$ & $^{11}$$M_{\rm ps}$ & $^{12}$$R_{\rm ps}$ \\
                & [Myr]         &               &       [$\Msun$]  &       [$\Msun$]  &                  &                 &                           & $[\mu m]$                &                       &        [$\Msun$]    &        [au]         \\
\hline \hline                                                                                                                                                                 
      {\tt F13} &       $86.6$  &       $10.2$  &       $2.72\E{6}$&       $3.00\E{5}$&       $-9.25$    &       $3.80$    &       $3.57\E{-8}$        &       $0.119$            &         5             &        0.0421       &        1.17         \\
      {\tt F50} &       $304$   &       $7.73$  &       $3.41\E{6}$&       $1.85\E{5}$&       $-7.94$    &       $5.06$    &       $4.52\E{-10}$       &       $0.012$            &         1             &        0.0294       &        1.62         \\
      {\tt F80} &       $564$   &       $6.01$  &       $7.47\E{6}$&       $2.73\E{5}$&       $-8.28$    &       $4.90$    &       $2.70\E{-8}$        &       $0.065$            &         1             &        0.0138       &        0.868        \\
\hline
\end{tabular}
\medskip \\
Note --- (1) ID of runs.
(2--3) time and redshift when SN shells return to original minihalos.
(4--5) dark matter and baryon mass of halos at $\tret$.
(6--7) carbon and iron abundance in recollapsing clouds.
(8--9) dust-to-gas mass ratio and characteristic grain radius of amorphous carbon (AC) 
after grain growth.
(10) number of fragments.
(11--12) mass and radius of the first hydrostatic core.
\end{minipage}
\end{table*}

\subsection{Recollapse of enriched clouds}

\begin{figure}
\includegraphics[width=0.99\columnwidth]{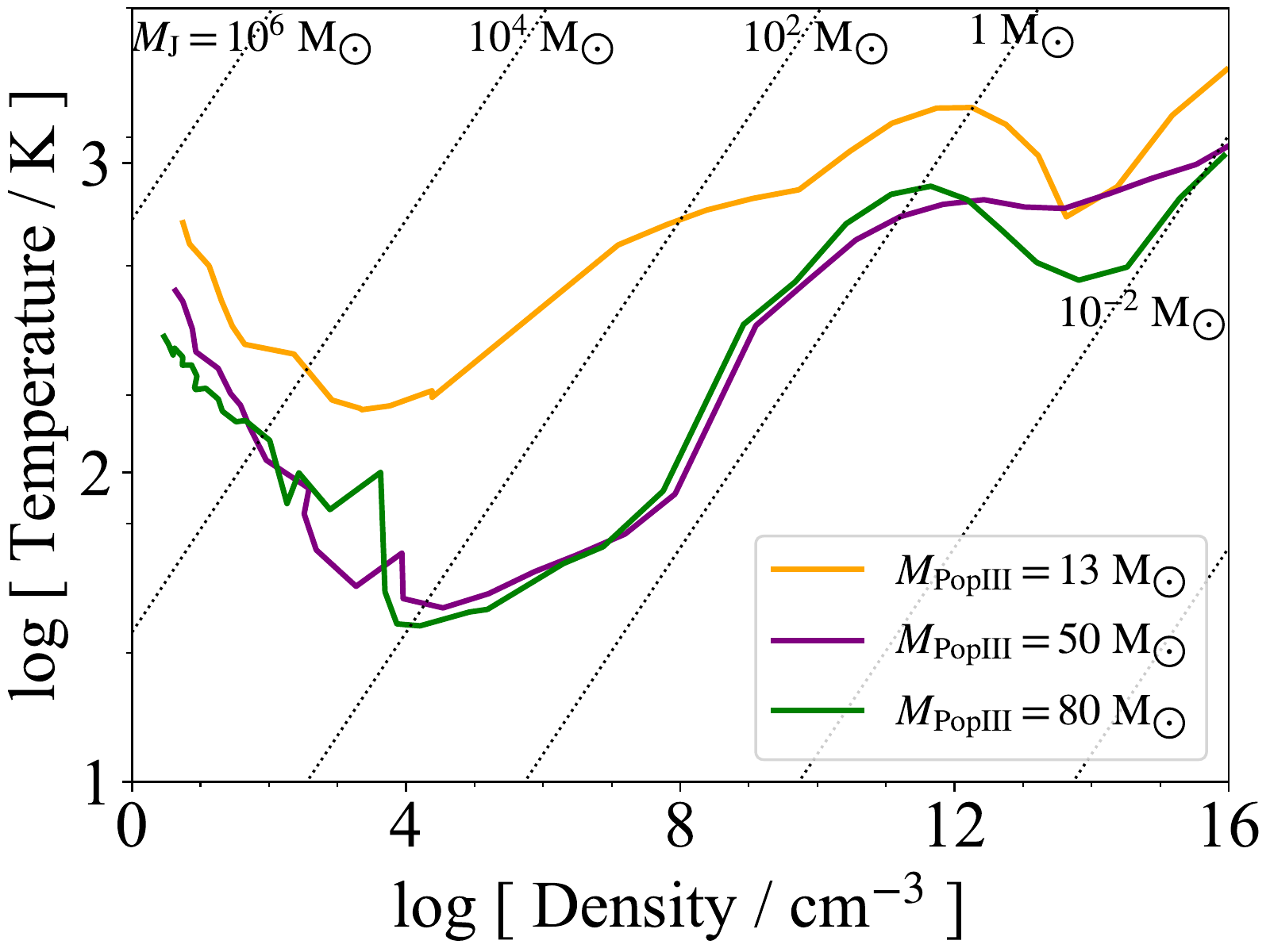}
\caption{
Temporal evolution of temperature of the cloud cores enriched by Pop III SNe with progenitor masses
$13 \ \Msun$ (orange), $50 \ \Msun$ (purple) and $80 \ \Msun$ (green).
We output snapshots at every time when the highest AMR level is refined and
plot the average density and temperature in the region with densities $> \nHmax /3$, where
$\nHmax$ is the maximum density.
}
\label{fig:nT}
\end{figure}

In this section, we investigate the chemothermal evolution of the cloud enriched by
faint SNe, comparing the results for the three progenitor models.
Fig. \ref{fig:nT} shows the thermal evolution of the clouds.
In the earlier stage of the evolution where the density of cloud center is 
below $\sim 10^{12} \ \percc$, molecular cooling is
dominant.
Since metal molecules can contribute to gas cooling for C abundances $\abA{C} \gtrsim 5$,
in Section \ref{sec:early_stage_of_collapse}, we discuss the thermal evolution in the early collapse stage,
focusing on the difference between the two cases with
\begin{itemize}
\item lower C abundance ($\abA{C} \sim 4$ for {\tt F13}),
\item higher C abundance ($\abA{C} \sim 5$ for {\tt F50} and {\tt F80}).
\end{itemize}
In the late collapse stage ($\nH \gtrsim 10^{12} \ \percc$), the thermal evolution 
mainly depends on the amount of AC grains through dust thermal emission cooling.
In Section \ref{sec:late_stage_of_collapse}, we compare the two cases with
\begin{itemize}
\item lower dust amount (${\cal D}_{\rm AC} \sim 10^{-10}$ for {\tt F50}),
\item higher dust amount (${\cal D}_{\rm AC} \sim 10^{-8}$ for {\tt F13} and {\tt F80}).
\end{itemize}

\begin{figure*}
\includegraphics[width=\textwidth]{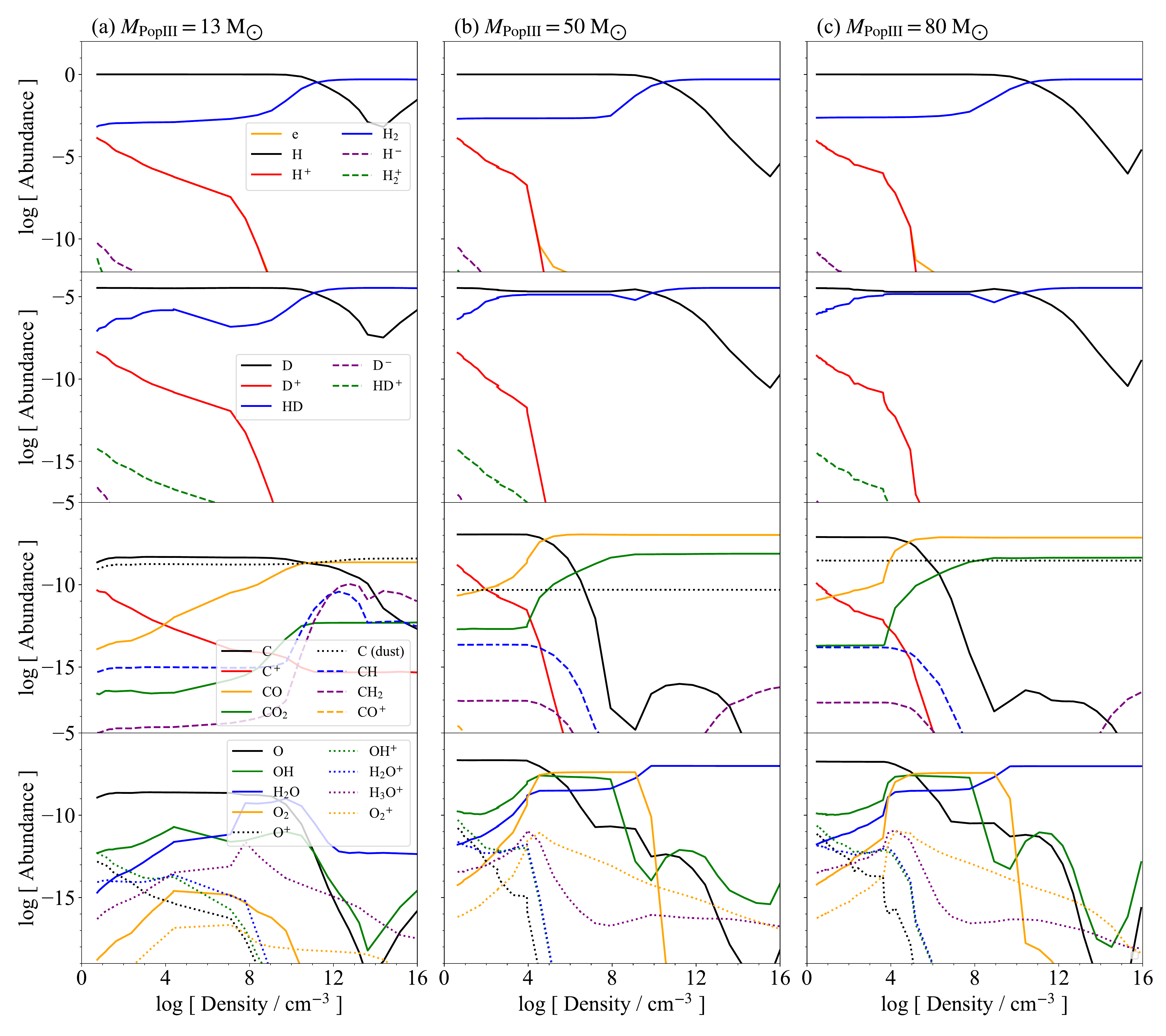}
\caption{
Evolution of the abundance of electrons, atoms, ions, molecules,
and grains in the cloud cores enriched by the Pop III SNe with progenitor masses
$13 \ \Msun$ (column a), $50 \ \Msun$ (column b) and $80 \ \Msun$ (column c)
as a function of density of the cloud cores.
}
\label{fig:ny}
\includegraphics[width=\textwidth]{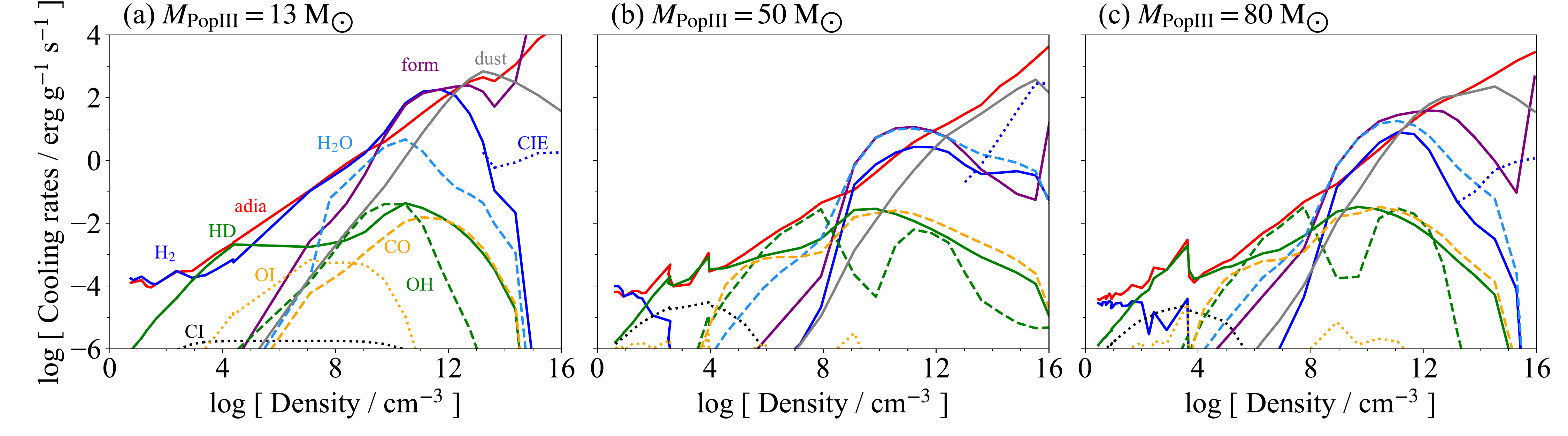}
\caption{
Evolution of cooling and heating rates of the cloud cores enriched by Pop III SNe with progenitor masses
(a) $13 \ \Msun$, (b) $50 \ \Msun$ and (c) $80 \ \Msun$ as a function of core density.
We plot the heating rates of
adiabatic gas compression (``adia''; red solid) and
H$_2$ formation (``form''; purple solid) and
radiative cooling rates of
H$_2$ (blue solid), HD (green solid), 
C {\sc i} (black dotted), O {\sc i} (orange dotted),
CO (orange dashed), OH (green dashed), H$_2$O (cyan dashed),
dust (grey solid), and CIE (blue dotted).
}
\label{fig:nL}
\end{figure*}

\begin{figure*}
\includegraphics[width=\textwidth]{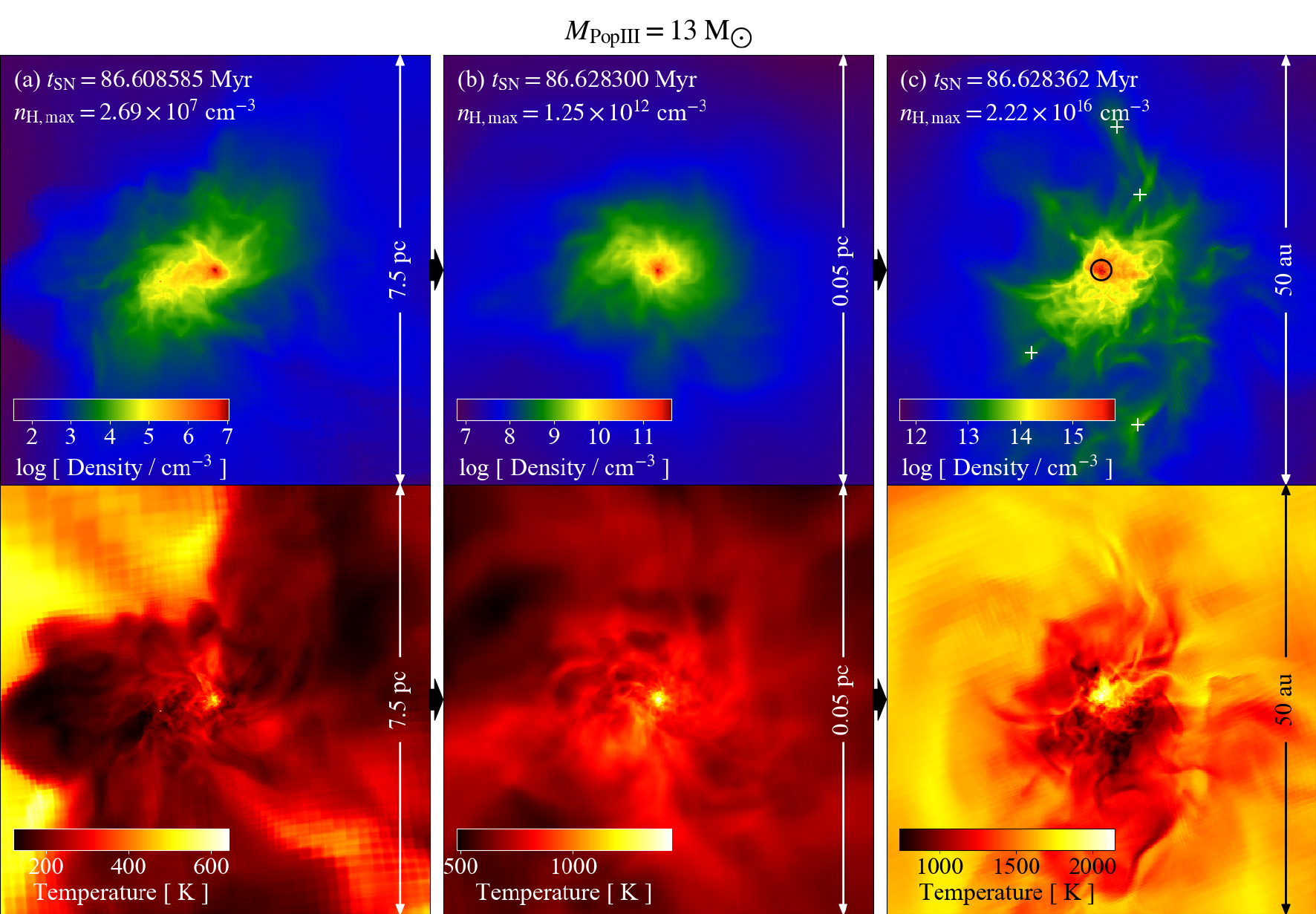}
\caption{
Density-weighted projection of density and temperature
in a box centered on the density maximum of the enriched clouds for
a progenitor mass $13 \ \Msun$.
We plot the snapshots at the time when the maximum density $\nHmax$ reaches
$\sim 10^7 \ \percc$ (column a), $10^{12} \ \percc$ (column b), and $10^{16} \ \percc$ (column c).
The circle and plus symbols are plotted on the density maxima of a hydrostatic core and 
clumps forming through fragmentation induced by dust thermal emission cooling, respectively.
The $x$ and $z$ axes of the windows are respectively parallel to the major and
minor axes of the momentum of inertia of the region with densities $> \nHmax /3$.
}
\label{fig:snapshots_col_M13}
\end{figure*}

\begin{figure*}
\includegraphics[width=\textwidth]{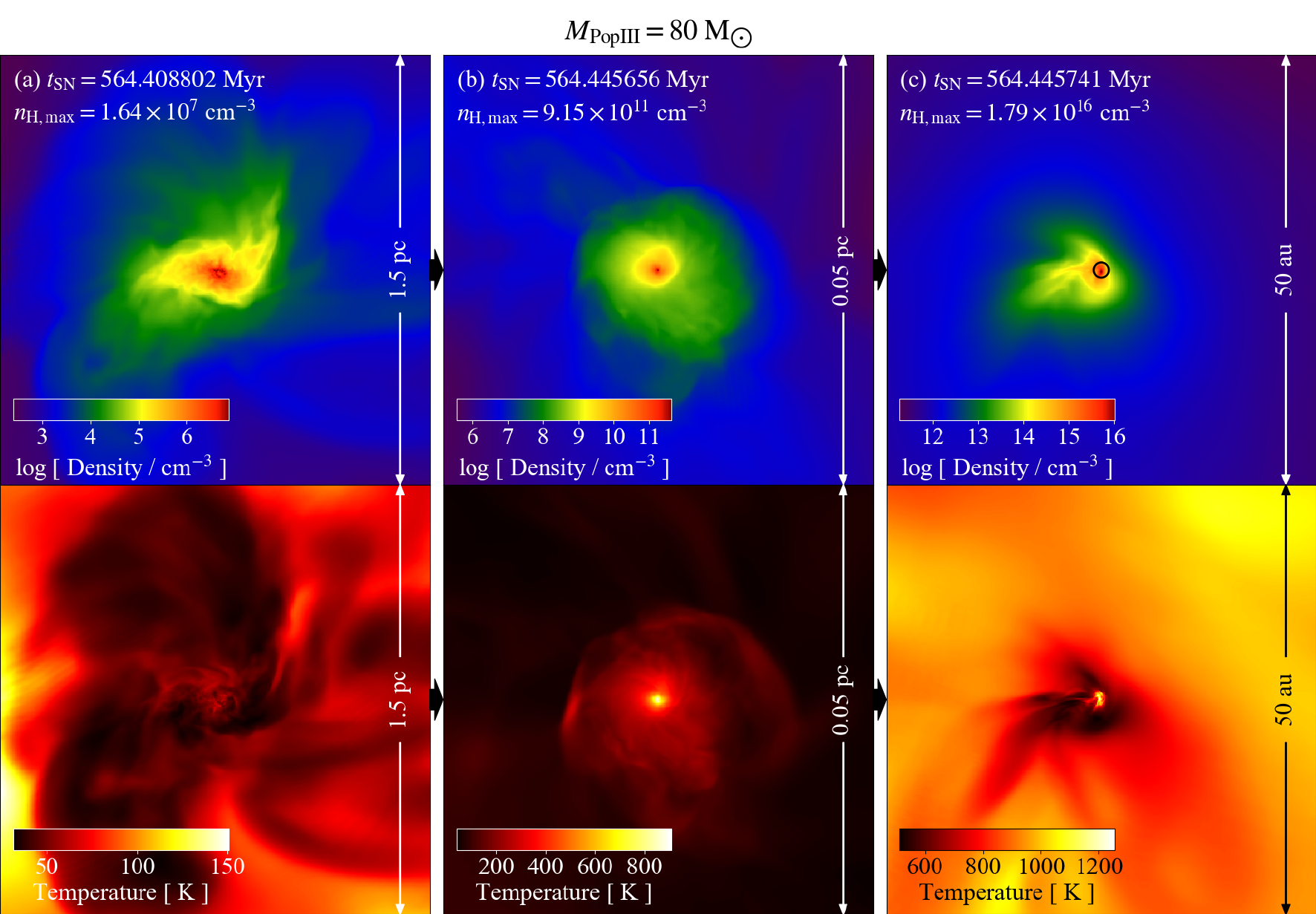}
\caption{
Same as Fig. \ref{fig:snapshots_col_M13} but for a progenitor mass $80 \ \Msun$.
In this run, fragmentation does not occur and thus only a hydrostatic core is shown by the circle.
}
\label{fig:snapshots_col_M80}
\end{figure*}

\begin{figure}
\includegraphics[width=\columnwidth]{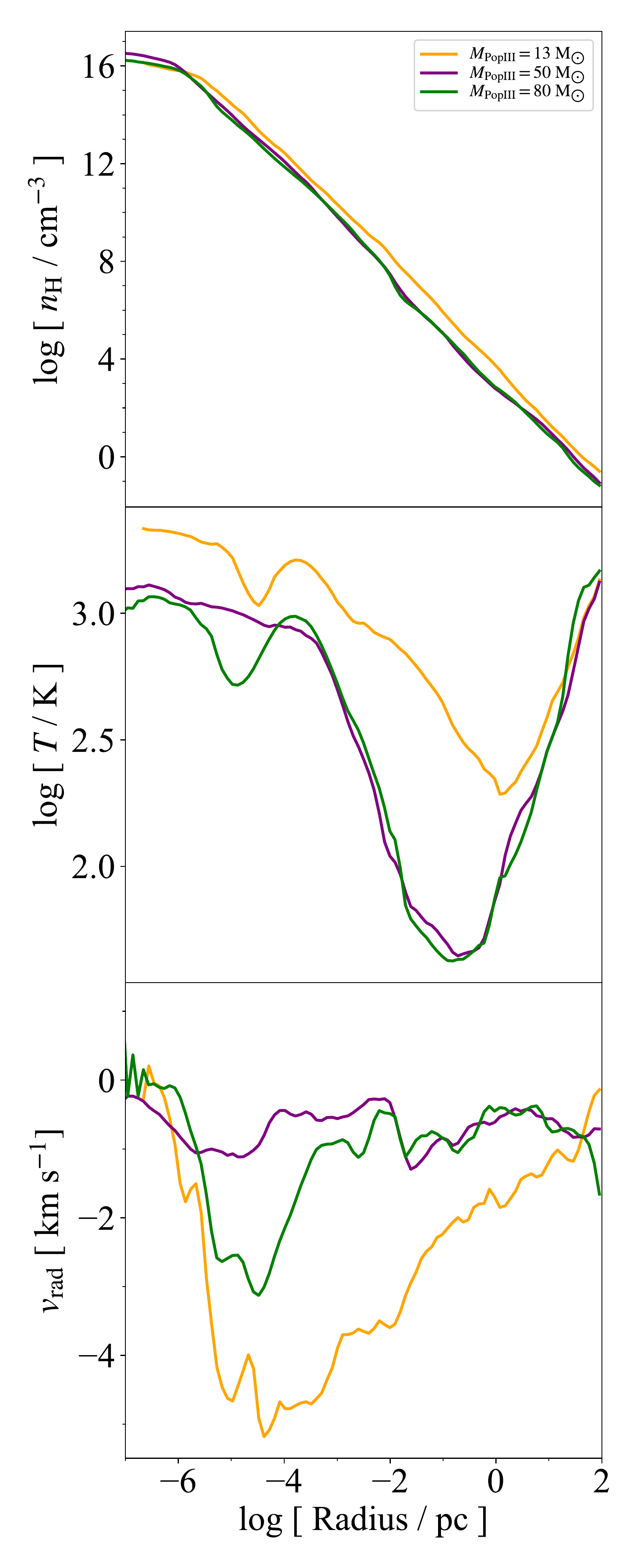}
\caption{Radial profile of density, temperature, and radial velocity at the time when we terminate our
simulations for progenitor masses $13 \ \Msun$ (orange), $50 \ \Msun$ (purple) and $80 \ \Msun$ (green).}
\label{fig:rv}
\end{figure}

\begin{figure*}
\includegraphics[width=\textwidth]{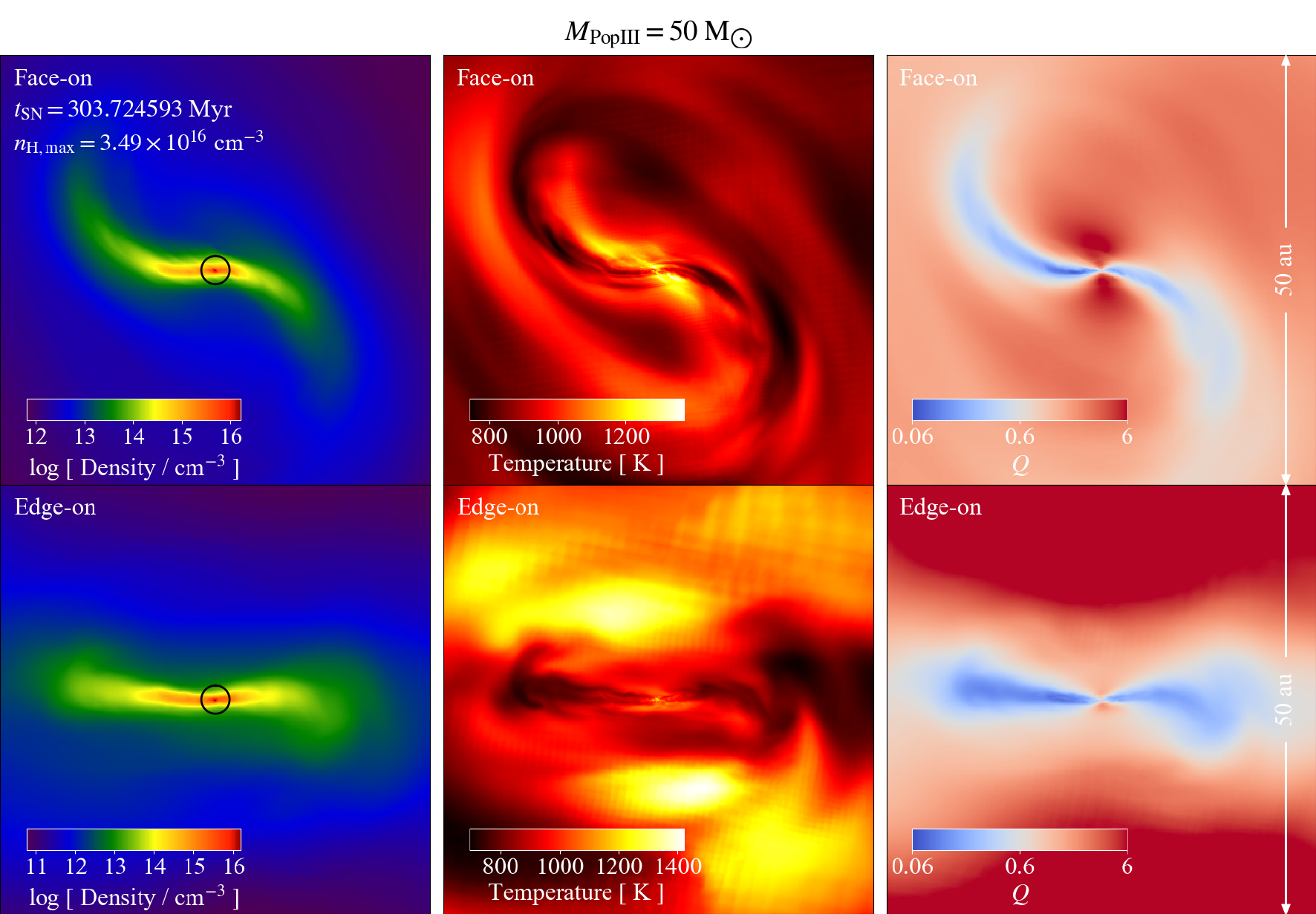}
\caption{
{\it Left to right}: density-weighted projection of density, temperature, and the Toomre $Q$ parameter
(see Section \ref{sec:growth_of_protostars}) 
in a box with a side 50 au centered on the density maximum of the enriched clouds for
a progenitor masses $50 \ \Msun$.
The top and bottom panels of each column show the disk from the face-on and edge-on view,
respectively.
The circle denotes the radius of a hydrostatic core forming in the central optically thick
region.
}
\label{fig:snapshots_fin_M50}
\end{figure*}

\subsubsection{Early stage of collapse: $\nH \lesssim 10^{12} \ \percc$}
\label{sec:early_stage_of_collapse}

Figs. \ref{fig:ny} and \ref{fig:nL} show the evolution
of the abundances of electrons, ions, molecules, and grains
and of cooling/heating rates, respectively,
as a function of the central density of the cloud core.
At densities $\nH \sim 1$--$10^2 \ \percc$, hydrogen molecules are the dominant coolant.
H$_2$ molecules form mainly through the H$^-$-process:
\begin{eqnarray}
{\rm H} + {\rm e^-} &\to & {\rm H^-} + \gamma , \nonumber \\ 
{\rm H^-} + {\rm H} &\to & {\rm H_2} + {\rm e^-}. \nonumber 
\end{eqnarray}
The grain surface reaction also contributes to molecular formation. 
The H$_2$ fraction relative to hydrogen nuclei becomes $\sim 10^{-3}$ for all models.
When the temperature drops below $150$ K, HD molecules form through the reactions
\begin{eqnarray}
{\rm H_2} + {\rm D} &\to & {\rm HD} + {\rm H} , \nonumber \\
{\rm H_2} + {\rm D^+} &\to & {\rm HD} + {\rm H^+}. \nonumber 
\end{eqnarray}
HD cooling becomes dominant at $\nH \sim 10^2$--$10^4 \ \percc$.
Columns (a) of Figs. \ref{fig:snapshots_col_M13} and \ref{fig:snapshots_col_M80} show
the morphology of the cloud core for {\tt F13} and {\tt F80} models, respectively.
We do not show the plot for {\tt F50} because the cloud morphology is similar to that for {\tt F80}
in the early collapse stage.
For all models HD cooling leads to a bar-mode instability
on a spherically symmetric core \citep{Lai00, Hanawa00}.
This cloud deformation is the precursor of fragmentation \citep{Tsuribe06}.
However, low-mass stars which can survive until the present day are unlikely to form through 
molecular cooling because the corresponding Jeans mass is still relatively large 
($\Mjeans \sim 10^2$--$10^4 \ \Msun$)
as indicated by the black dotted lines in Fig. \ref{fig:nT}.

At densities $\nH \sim 10^4$--$10^8 \ \percc$, the temperature is smaller ($30$ K)
for {\tt F50} and {\tt F80} than for {\tt F13} ($150$ K)
because OH cooling becomes dominant in the former case.
OH and O$_2$ molecules form mainly through the reactions
\begin{equation}
{\rm O} + {\rm H} \to  {\rm OH} + \gamma . \nonumber
\end{equation}
at a density $\sim 10^4 \ \percc$.
For {\tt F13}, OH cooling is less efficient because the
O abundance ($\abH{O} = -5.33$) is smaller than for the other models
($\abH{O} = -3.34$ and $-3.44$ for {\tt F50} and {\tt F80}, respectively).
Further, O is rapidly depleted into CO molecules because
C has a larger abundance with respect over O (${\rm [C/O]} = +0.691$) compared to the other models
(${\rm [C/O]} = -0.0269$ and $-0.0855$ for {\tt F50} and {\tt F80}).

At densities $\nH \sim 10^8$--$10^{12} \ \percc$, the temperature increases rapidly due
to gas heating along with H$_2$ formation through three-body reactions
\begin{eqnarray}
{\rm H} + {\rm H} + {\rm H} &\to & {\rm H_2} + {\rm H} , \nonumber \\
{\rm H} + {\rm H} + {\rm H_2} &\to & {\rm H_2} + {\rm H_2} . \nonumber 
\end{eqnarray}
The temperature increase is more significant in models {\tt F50} and {\tt F80}
than in model {\tt F13} 
because the temperature just before H$_2$ formation heating becomes dominant
is smaller ($\sim 100$ K) owing to cooling from OH as well as H$_2$O molecules 
forming through the reaction
\begin{equation}
{\rm OH} + {\rm H_2} \to {\rm H_2O} + {\rm H}. \nonumber
\end{equation}
This rapid gas heating stabilizes the cloud against the deformation of the cloud core.
A hot spherical hydrostatic core defined by accretion shocks 
forms at a radius $\sim 10^{-2}$ pc for {\tt F80} (Fig. \ref{fig:snapshots_col_M80}b).
Whereas, density fluctuations remain within this radius scale through 
the moderate gas heating for {\tt F13} (Fig. \ref{fig:snapshots_col_M13}b).
Fig. \ref{fig:rv} shows the radial profile of density, temperature, and radial velocity.
The velocity increases inward by $\sim 1 \ \kmpers$ at a radius
$\sim 10^{-2}$ pc where temperature rapidly increases for {\tt F50} and {\tt F80}
while the velocity smoothly decreases for {\tt F13}.

\subsubsection{Late stage of collapse: $\nH \gtrsim 10^{12} \ \percc$}
\label{sec:late_stage_of_collapse}

For {\tt F13}, AC grains grow in size through accretion of gas phase C atoms.
Fig. \ref{fig:dist} shows the size distribution of AC grains before
(dashed curves) and after (solid curves) grain growth.
The size distribution shifts toward the larger radii by $\Delta r = 0.02 \ \um$.
Fig. \ref{fig:ny}a shows that the abundance of C atoms locked up into AC grains (black dotted curve)
slightly increases at a density $\nH \sim 10^{13} \ \percc$.
The dust-to-gas mass ratio increases from $1.49\E{-8}$ to $3.57\E{-8}$ 
(Tables \ref{tab:SN} and \ref{tab:metallicity}) and the grain cooling rate is enhanced.
For the other models, grain growth is not effective because 
C atoms are mostly oxidized into CO molecules at densities $\nH \sim 10^5 \ \percc$
due to the larger O/C abundance ratio.

At densities $\nH = 10^{12}$--$10^{14} \ \percc$, while the temperature evolves nearly isothermally 
for the model with low dust content ({\tt F50}),
the temperature decreases due to cooling by dust thermal emission for the models with
high dust content ({\tt F13} and {\tt F80}).
The condition where the dust cooling rate exceeds the adiabatic compressional heating
(the main heating source of the clouds) at a given density $\nH$ and temperature $T$ is
\begin{eqnarray}
\frac{{\cal D}_{\rm AC}}{r_{\rm AC, cool}} &>& 1.86\E{-7} \ \um ^{-1}
\left( \frac{T}{500 \ {\rm K}} \right) ^{-1/2} \nonumber \\ 
&&\times
\left( \frac{\nH}{10^{13} \ \percc} \right) ^{-1/2} .
\label{eq:crit}
\end{eqnarray}
\citep{Schneider12, Chiaki15}.
Assuming the values ${\cal D}_{\rm AC}$ and $r_{\rm AC, cool}$ achieved after grain growth (Table \ref{tab:metallicity}), 
we find that, for models {\tt F13} and {\tt F80},
${\cal D}_{\rm AC}$ / $r_{\rm AC, cool} = 3.00\E{-7} \ \um ^{-1}$ and $4.15\E{-7} \ \um ^{-1}$, 
respectively, above the critical value
while, for {\tt F50} $3.77\E{-8} \ \um ^{-1}$, below the critical value.

Finally, in the region with densities $ > 10^{13} \ \percc$,
since the grains become thermally coupled with the gas, the temperature starts to increase again.
Then, the gas becomes optically thick to continuum radiation.
The cloud core evolves nearly adiabatically and a hydrostatic core, which
can not contract further, forms as we can see at a radius 
$\sim 1$ au ($\sim 10^{-6}$ pc) in Fig. \ref{fig:rv} \citep{Larson69, Penston69}.

\subsection{Fragmentation of enriched clouds}

Whether fragmentation occurs in the later stage of collapse or not 
depends on the different thermal evolution among the three progenitor models. 
For {\tt F50}, cloud fragmentation does not occur because of inefficient dust cooling (Eq. \ref{eq:crit}).
Fig. \ref{fig:snapshots_fin_M50} shows the snapshot of the cloud core in the later collapse stage for {\tt F50}.
A disk structure forms around a central hydrostatic core (circle)
as for primordial clouds \citep{Clark11, Greif12, Hirano17} because
the temperature evolution in the later stage is similar to the primordial clouds
as H$_2$ cooling is dominant (Fig. \ref{fig:nL}b).
In Fig. \ref{fig:snapshots_fin_M50}, we show the disk from the face-on and edge-on views.
The disk is gravitationally unstable and dense spiral arms form.
The arms might fragment in the further evolution of the accretion disk
as we discuss in Section \ref{sec:discussion}.

For higher dust content, although dust cooling becomes effective in both runs,
the cloud core fragments into isolated filamentary structures for {\tt F13} 
(Fig. \ref{fig:snapshots_col_M13}c) while
fragmentation does not occur for {\tt F80} (Fig. \ref{fig:snapshots_col_M80}c).
In the latter case, only short filaments connected with the central core appear.
For the filaments to fragment, their aspect ratio should grow up to 20--30
\citep{Tsuribe06, Chiaki16}.
In this model, rapid gas heating along with H$_2$ molecular 
formation has made the cloud stable against the deformation 
at densities $\nH = 10^8$--$10^{12} \ \percc$ before dust cooling occurs. 
Therefore, the timescale for deformation of the spherical cloud core to sufficiently
thin filaments is longer than the dynamical timescale of the cloud while
the opposite is true for {\tt F13} model.
As seen in Section \ref{sec:early_stage_of_collapse}, this difference of heating rate is due to the different 
cooling efficiency of metal molecules (OH/H$_2$O) 
before the onset of H$_2$ formation heating.

Our detailed chemistry/cooling model shows that the fragmentation properties of the clouds enriched by faint SNe
depend on multiple cooling/heating processes (1) dust thermal emission cooling,
(2) H$_2$ formation heating, and (3) OH/H$_2$O molecular cooling \citep[see also][]{Chiaki16}.

\begin{figure*}
\includegraphics[width=12cm]{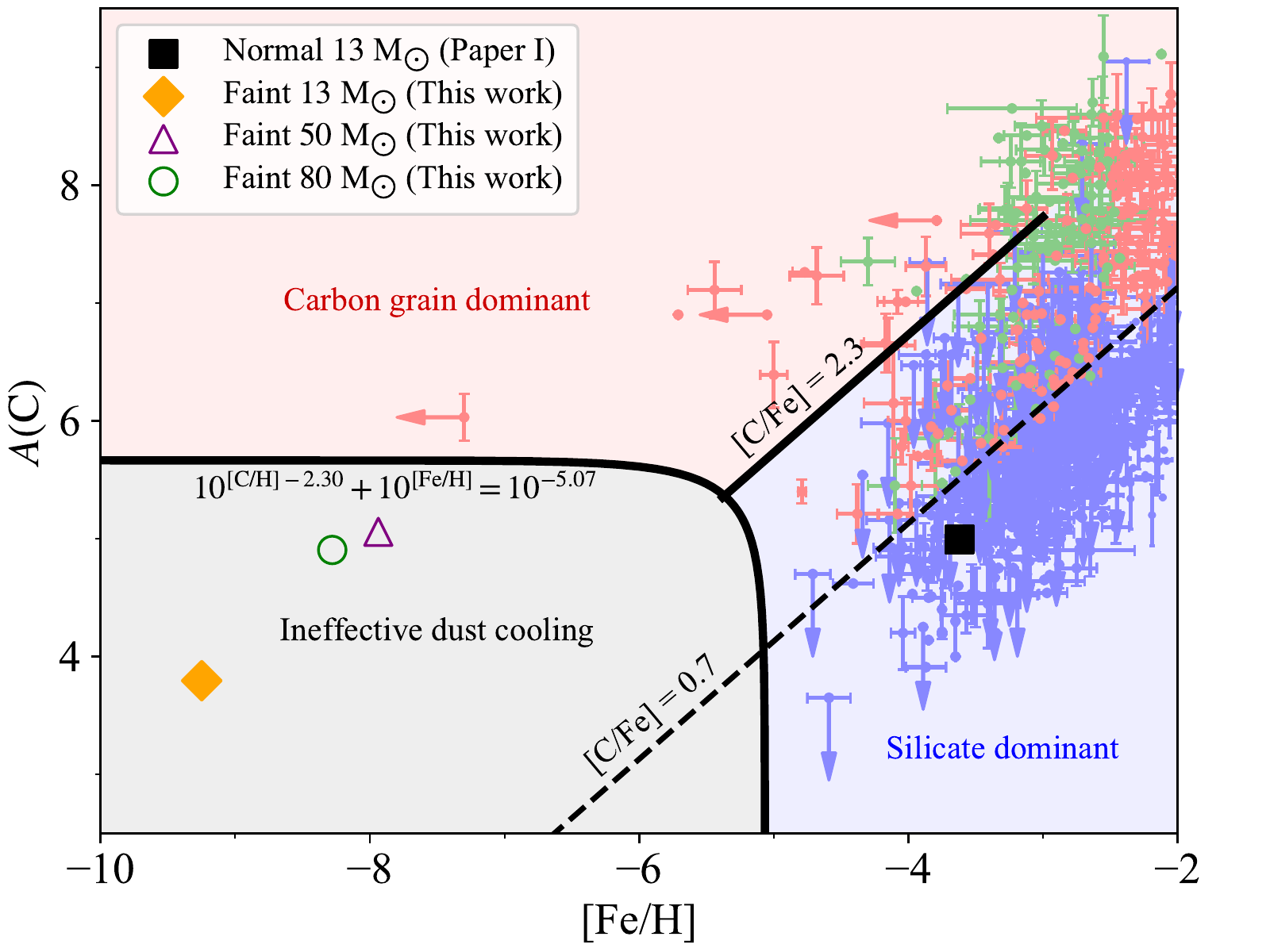}
\caption{
Elemental abundances of clouds enriched by faint SNe for progenitor masses 
$\MPopIII = 13 \ \Msun$ (orange diamond), $50 \ \Msun$ (purple triangle), and $80 \ \Msun$ (green circle)
on the $\abA{C}$-$\abH{Fe}$ plane.
The black square shows the C and Fe abundances of the clouds enriched by a normal CCSN
with a $\MPopIII = 13 \ \Msun$ from our previous simulation \citepalias{Chiaki19}.
The filled and open symbols indicate the runs where cloud fragmentation occurs or not, respectively.
The dots show the abundances of observed stars
retrieved from the SAGA database \citep{Suda08}:
{\it blue dots} C-normal stars ($\abFe{C} < 0.7$) or stars whose C abundances are not constrained, 
{\it green dots} CEMP-s stars ($\abFe{C} > 0.7$ and $\abFe{Ba} > 0.5$), and
{\it red dots} CEMP-no stars ($\abFe{C} > 0.7$ and $\abFe{Ba} < 0.5$) or
CEMP stars whose Ba abundances are not constrained.
We also plot the criterion for CEMP stars \citep[$\abFe{C} = 0.7$;][]{Aoki07}.
The black solid line shows the critical condition defining the carbon grain dominant 
region (red shaded) and silicate dominant region (blue shaded).
In the grey shaded region below the black solid curve, 
no stars have been identified, which may suggest that
dust cooling is ineffective in their parent clouds and low-mass stars hardly form
\citepalias{Chiaki17}.
}
\label{fig:C_Fe}
\end{figure*}

More quantitatively, we count the number and mass of clumps 
within 100 au from the maximum density with
a clump-finding algorithm.
This identifies isolated clumps by using iso-density contours \citep{Smith09, yt}.
If a candidate clump does not meet the user-defined validators,
it is merged with another clump or discarded.
In this work, we impose the following three criteria:
\begin{enumerate}
\item the minimum number of cells in a clump is $20$, 
\item the minimum density of the cell in a clump is $n_{\rm H, th} = 10^{13} \ \percc$,
\item the minimum mass of a clump is $\abM{th} = 10^{-3} \ \Msun$ ($\sim 1$ Jupiter mass).
\end{enumerate}
In the criteria (ii), we choose the threshold density $n_{\rm H, th} = 10^{13} \ \percc$,
where dust cooling becomes dominant, to select clumps formed through dust-induced fragmentation.
Here we do not require the clumps to be gravitationally bound.
For the most massive clump, which has developed a hydrostatic core,
we estimate the mass $M_{\rm ps}$ and radius $R_{\rm ps}$ above
which the gravitational energy $3 G M_{\rm ps}^2 / 5R_{\rm ps}$ 
exceeds the total thermal energy.
We hereafter call the primary clump as ``protostar''.
Since the computational time is set by the short dynamical time of the 
dense protostar, when we terminate the simulation the secondary clumps are 
still in the early phase of their gravitational growth.
Therefore, in this work we just show their mass derived from the clump-finding algorithm.
We hereafter call the secondary clumps as just ``clumps''.

For {\tt F13}, the number $N_{\rm frag}$ of identified clumps is five (Table \ref{tab:metallicity}).
The circle and plus symbols in Fig. \ref{fig:snapshots_col_M13}c depict the distribution
of the primary protostar and the secondary clumps, respectively.
The mass and radius of the protostar are ($M_{\rm ps}$, $R_{\rm ps}$) = ($0.0421 \ \Msun$, $1.17$ au).
The masses of the four clumps are $1.06\E{-2}$, $2.35\E{-3}$, $1.55\E{-3}$, and $1.21\E{-3} \ \Msun$.
The separation of clumps ($\sim 10$ au) and mass of the protostar
are comparable to the local Jeans length and mass for $\nH = 10^{14} \ \percc$
and $T = 1000$ K, which indicates that the fragmentation is induced by
dust cooling.
For {\tt F50} and {\tt F80}, the clump-finding algorithm can not identify any secondary
clumps ($N_{\rm frag} = 1$).
The mass and radius of the protostar are
($0.0294 \ \Msun$, $1.62$ au) and
($0.0138 \ \Msun$, $0.868$ au)
for {\tt F50} and {\tt F80}, respectively.

Although these protostars and clumps will gain mass by accreting the ambient gas in the subsequent evolution 
until they reach the main-sequence as discussed in Section \ref{sec:discussion},
dust-induced low-mass fragments can form in model {\tt F13}.

\section{Discussion}
\label{sec:discussion}

\subsection{Prediction of observations}

We have so far presented the results of our simulation of metal enrichment from faint SNe and
the fragmentation properties of the enriched clouds.
If the protostars in the clouds grow to low-mass stars and the MH is accreted into the Milky 
Way halo or Local Group dwarf galaxies through DM halo mergers,
they can be observed as EMP stars.
In this section, we discuss their observability.  
The colored symbols in Fig. \ref{fig:C_Fe} depict the distribution of elemental abundances of 
the enriched clouds on the $\abA{C}$-$\abH{Fe}$ diagram used in \citet{Yoon16}.
We also plot the abundances of protostars forming in the cloud enriched from a normal CCSN
with a $\MPopIII = 13 \ \Msun$ (black square) 
in the simulation of our previous study \citepalias{Chiaki19}, hereafter called the run {\tt C13}.
The filled and open symbols indicate the runs where fragmentation occurs and not, respectively.

For comparison, we plot the abundances of observed EMP stars from the SAGA database \citep{Suda08}.
The blue dots represent the C-normal stars with $\abFe{C} < 0.7$ or
stars whose C abundances are not constrained. 
The other colors represent C-enhanced stars with $\abFe{C} > 0.7$, following the
definition of \citet{Aoki07}.
The green and grey dots represent CEMP-s with $\abFe{Ba} > 0.5$ and
CEMP-no with $\abFe{Ba} < 0.5$ or stars whose Ba abundances are not constrained, respectively.
As \citetalias{Chiaki17} found, no stars have been observed
in the grey shaded area bounded by the black solid curve.

From a comparison of our simulation results with the observational data,
we can see that
the abundances for {\tt C13} ($\abH{Fe}$, $\abA{C}$) = ($-3.62$, $4.99$)
are in the range of observed C-normal stars ($\abH{Fe} > -5$) and
can explain the formation of C-normal stars.
Whereas, the abundances for {\tt F13}, {\tt F50}, and {\tt F80} ($\abH{Fe} = [-9.25:-7.94]$ and
$\abA{C} = [3.80:5.06]$) are below the range of observed CEMP Group III stars ($\abA{C} \gtrsim 6$)
although the abundance ratio $\abFe{C} = 4.57$--$4.75$ is consistent
with the observed values.

We now discuss the reason why the elemental abundances for the faint SN models
are systematically smaller than those for the normal CCSN model.
The elemental abundances of recollapsing clouds depend on the following three parameters
\citep[][\citetalias{Chiaki19}]{Ritter15}.
First, the mass of an element M ejected from SNe, $\abM{M}$, is an important value.
Second, the fraction $\ffb$ of metals that return to the MH determines the elemental abundances
of the enriched clouds.
We can estimate this as the fraction of solid angles where
metal dispersion is interrupted on the contact surfaces between the SN shell
and cosmic filaments or interstellar gas clumps.
Finally, the dilution efficiency of the metals is important, i.e., the mass of pristine gas with which the metals
are mixed. 
The simplest estimate is to use the pristine gas mass within the virial radius, similar to 
semi-analytic works \citep[e.g.,][]{deBennassuti17, Hartwig18}.
Since the distribution of elements is not uniform within the virial
radius of the MH, we consider only the dense central cloud where the elemental abundances
are uniform (Fig. \ref{fig:rZ}).
In this work, we define the pristine gas mass as the Jeans mass $\abM{cloud}$,
at the density $10^{3} \ \percc$ (see the black dotted lines in Fig. \ref{fig:nT}).
With these three parameters ($\abM{M}$, $\ffb$, and $\abM{cloud}$),
we can estimate C and Fe abundances in the enriched cloud to be
\begin{eqnarray}
A ({\rm C}) &=& 12 + \log \left( \frac{\ffb \abM{C} / \muX{C}}{\XH \abM{cloud}} \right), \\
{\rm [Fe/H]} &=& 12 + \log \left( \frac{\ffb \abM{Fe} / \muX{Fe}}{\XH \abM{cloud}} \right) - \abAsun{Fe},
\end{eqnarray}
where $\muX{C} = 12$ and $\muX{Fe} = 56$ are the molecular masses of C and Fe, respectively.

For {\tt C13}, the ejected mass of C and Fe is ($\abM{C}$, $\abM{Fe}$) = ($0.5 \ \Msun$, $0.2 \ \Msun$).
The metal dispersion is interrupted by a clump with a distance $D_{\rm cl} = 50$ pc 
and radius $R_{\rm cl} = 6$ pc.
Therefore, $\ffb = \pi R_{\rm cl}^2 / 4\pi D_{\rm cl}^2 \sim 0.004$.
With a cloud mass $\abM{cloud} = 2000 \ \Msun$, we can estimate the abundances as 
($\abH{Fe}$, $\abA{C}$) = ($-3.6$, $5.0$),
consistent with the simulation result.\footnote{Compared to the estimation in
\citetalias{Chiaki19}, we improve the approximation accuracy, and thus the estimation is 
slightly changed.}
Even with the small $\ffb$, we can explain the formation of C-normal EMP stars.
For {\tt F13}, the mass of ejected metals are ($\abM{C}$, $\abM{Fe}$) = ($0.08 \ \Msun$, $10^{-6} \ \Msun$).
Since the density structure of ISM is similar to that for {\tt C13},
the return fraction is $\ffb \sim 0.004$.
With the cloud mass $\abM{cloud} = 4000 \ \Msun$,
($\abH{Fe}$, $\abA{C}$) = ($-9.2$, $3.9$).
Compared to {\tt C13}, the ejected C mass is smaller by a factor of six.
Also, the cloud mass is larger because the gas temperature is higher due to the smaller metallicity.
Consequently, the solid angle where metal dispersion is interrupted becomes smaller.

For {\tt F50} and {\tt F80}, the metal yield is ($\abM{C}$, $\abM{Fe}$) = ($1 \ \Msun$, $10^{-5} \ \Msun$).
The radius of the interacting region of the shell with the ambient clump is 
larger $D_{\rm cl} = 100$ pc than for {\tt F13}
because filaments and clumps within $R_{\rm cl}$ are photoevapolated by the strong UV emission.
It takes longer time until the SN shells return and the clump has gravitationally grown
to $R_{\rm cl} = 7$ pc.
With $\ffb = \pi R_{\rm cl}^2 / 4\pi D_{\rm cl} \sim 0.001$ and $\abM{cloud} = 1000 \ \Msun$,
($\abH{Fe}$, $\abA{C}$) = ($-8.0$, $5.1$), which is consistent with our simulation results.
Compared to {\tt C13}, the ejected C mass is larger by a factor of three.
However, $\ffb$ is smaller by a factor of three because the radius of the region within which
metals are mixed is larger 
with larger explosion energies ($\Esn = 2.6$--$5.2\E{51}$ erg).
Consequently, $\abA{C}$ in the enriched clouds becomes comparable to that for {\tt C13}.
The value is smaller than $\abA{C}$ of the observed CEMP Group III stars.

For the MH and Pop III faint SNe investigated here, 
the enrichment from a single faint SNe can hardly explain
the formation of CEMP stars with the observed metallicity range.
We discuss their formation paths in general cases in Section \ref{sec:general_initial_conditions}.

We also find that cloud fragmentation occurs through dust cooling for {\tt F13}
even with the lowest metallicity ($\abH{Fe} \sim -9$).
This is because the condensation efficiency of C into AC grains is large
($f_{\rm AC, C} = 0.262$) and 
the cloud avoids rapid H$_2$ formation heating (Section \ref{sec:early_stage_of_collapse}).
This indicates that low-mass stars with a metallicity $\abH{Fe} \sim -9$ can form
but the abundance is below the region where CEMP stars have been observed so far
(grey shaded region in Fig. \ref{fig:C_Fe}).

\citetalias{Chiaki17} estimated the effective grain radius of AC to be
$r_{\rm AC, cool} / f_{\rm AC, C} \sim 10 \ \um$
at which the dust cooling rate overcomes the compressional heating rate (Eq. \ref{eq:crit}) 
above the critical C abundance $\abAcr{C} \sim 6$.
The dust model of \citetalias{Marassi14} used in this work
suggests the formation of smaller grains with 
$r_{\rm AC, cool} / f_{\rm AC, C} = 0.454 \ \um$ for {\tt F13}
in faint SNe.
Grains smaller than $10 \ \um$ can induce cloud fragmentation for C abundances
lower than $\abAcr{C} \sim 6$.

A semi-analytic study of galactic chemical evolution including the contributions from faint SNe 
\citep{Komiya20} gives a consistent result with our simulations.
They point out that faint SNe might overproduce stars with significantly small 
metallicities ($\abH{Fe} < -7$) compared to the metallicity distribution function (MDF) 
of observed stars because these SNe produce smaller mass of metals than normal 
CCSNe \citep[see also][]{deBennassuti14, deBennassuti17}.
Unfortunately, the MDF in the low metallicity end still has a large uncertainty because the number
of observed stars with $\abH{Fe} < -4.5$ is small ($\sim 10$).
Future surveys and spectroscopic studies with 
Subaru/PFS \citep{Takada14} and the Thirty-Meter Telescope (TMT) \citep{Skidmore15}
will probe the chemical enrichment process in the early Universe.

\subsection{Caveats}

\subsubsection{General initial conditions}
\label{sec:general_initial_conditions}

In this work, we have investigated the metal enrichment from faint SNe 
with progenitor masses $\MPopIII = 13$, $50$, and $80 \ \Msun$
in a MH with a $\Mhalo = 2.1\E{6} \ \Msun$.
The MH is relatively high-mass, compared to
the mass range of MHs derived from large-volume cosmological simulations, 
$2\E{5}$--$3\E{6} \ \Msun$ \citep{Susa14, Hirano14, Hirano15}.
\citet{Chiaki18} investigated CEMP star formation for different initial conditions
with $\Mhalo = 3\E{5}$--$3\E{6} \ \Msun$ and $\MPopIII = 20$--$30 \ \Msun$,
and found that the elemental abundances in recollapsing clouds are larger for lower-mass MHs.
For a high-mass MH ($3\E{6} \ \Msun$), the SN shell becomes gravitationally unstable during its expansion
because the \HII \ region does not expand outside the virial radius, 
and the shell rapidly loses its thermal energy once it encounters the neutral medium.
The SN shell enriches a nearby cloud and collapses away from the explosion site, offset from the halo center. 
A smaller fraction of metals mixes into the cloud that has already begun to collapse. 
Consequently, the elemental abundances are smaller than the low-mass case ($3\E{5} \ \Msun$), 
where the explosion disrupts the MH and recollapses into the halo center.
For low-mass MHs, the iron abundance can be estimated to be in the range $ \abH{Fe} = [-8.63: -4.34]$, and thus 
the formation of CEMP stars can be explained by the enrichment from a single faint SN
in several cases.
In \citet{Chiaki18}, the resolution was limited ($\sim 1$ pc) compared to this work,
and the formation of dust grains was not considered.
We will investigate CEMP star formation for a wide range of MH and Pop III stellar masses with
a resolution similar to the present study in large cosmological simulations in future works.

\subsubsection{Multiple source of metals}
In this work, we restrict our focus on Pop III star formation in a single MH.
There are several MHs around the central MH within 1 kpc (Fig. \ref{fig:snapshots_ini})
and they might contribute to their mutual enrichment \citep{Smith15}.
For {\tt F80}, it takes $500$ Myr until the metals from the central MH return
and during this time interval the metals ejected by other MHs may possibly enrich the MH.
The effect of multi-enrichment will be to create a superposition of elemental abundances having
multiple SN progenitors \citep{Salvadori07, Salvadori10, deBennassuti15, deBennassuti17, Hartwig18}.
However, in this work we consider the formation of a single Pop III star in the MH.
Numerical simulations of primordial clouds suggest that the 
stellar cluster formation through disk fragmentation may occur
\citep{Turk09, Clark11, Greif12}.
A cosmological simulation has also indicated that multiple Pop III stars form
in each MH \citep{Skinner20}.
In the case of multiple SNe in a MH,
\citet{Ritter15} found that the gas in the MH is evacuated by 
the first SN blast wave but a non-negligible fraction ($\sim 0.3$) of metals ejected from
the secondary SNe contribute to the enrichment of the MH. 

In galactic archaeology, the number of progenitors of EMP stars is
important value as
the mass and explosion energy of the Pop III stars can
be directly traced back from the elemental abundances of singly enriched stars 
\citep{Ishigaki18, Hartwig18, Choplin19}.
In their semi-analytic studies, \citet{deBennassuti17} predict that stars with metallicities 
$\abH{Fe} \lesssim -5$ can be
the very second generation of stars, which is the metallicity range
that we have investigated.
In order to study the formation of stars with larger metallicities,
we will investigate multiple star formation in each MH and mutual metal enrichment
between close MHs in cosmological simulations to see the frequency of multi-enrichment
in forthcoming papers.

\subsubsection{Growth of protostars}
\label{sec:growth_of_protostars}

When we terminate the simulations, protostars have formed in the optically thick region.
Their mass is $\sim 0.01 \ \Msun$, comparable to the Jeans mass with density
$\sim 10^{14} \ \percc$ and temperature $\sim 1000$ K. 
We should note that, until the protostars reach the zero-age main-sequence, 
their masses grow through the accretion of the ambient materials \citep{Hosokawa11, Hirano14}.
For {\tt F13}, if the protostars remain low-mass, with a mass less than $0.8 \ \Msun$ 
despite competitive accretion among the fragments, 
the stars will be observed in the present day.
For {\tt F50} and {\tt F80}, gas accretion is partly halted by ionization
feedback from the protostar but the protostar can grow up to $\sim 100 \ \Msun$
with a sufficiently large accretion rate $\sim 10^{-3} \ \Msunyr$ (Fukushima et al. in prep.).
The massive star will explode as a SN and contribute to the enrichment of 
the succeeding generation of stars.

We should also note that the accretion disks forming around the protostars are 
in most cases gravitationally unstable to fragment \citep{Turk09, Clark11, Greif12}.
This {\it disk fragmentation} can occur even in the primordial clouds.
This fragmentation mode is distinctive from 
{\it cloud fragmentation} induced by radiative cooling \citep{Chiaki16}.
Although the critical conditions for disk fragmentation are still debated
\citep{Chon19, Liao19, Inoue20}, \citet{Takahashi16} find that a spiral arm structure
developing in a disk can fragment when its minimum Toomre 
$Q = \cs \Omega _{\rm epi} / \pi G \Sigma$ becomes less than 0.6, 
where $\Omega _{\rm epi}$ is the epicyclic frequency and
$\Sigma$ is the column density in the perpendicular direction of the disk.
For {\tt F50}, spiral arms developed (Fig. \ref{fig:snapshots_fin_M50}).
We estimate the Toomre $Q$ parameter by approximating $\Omega _{\rm epi} = 2\Omega$, where $\Omega$
is the gas angular momentum, and we find that $Q$ in the spiral arms is below 0.6 
(blue colored region in the upper right panel of Fig. \ref{fig:snapshots_fin_M50}),
which suggests that the arms will fragment.
However, since the gas accretion rate onto the protostar is higher 
($\sim \cs ^3 /G \sim 3\E{-3} \ \Msunyr$)
than in present-day star-forming regions ($\sim 10^{-5} \ \Msunyr$),
disk fragmentation might result in the formation of a massive star cluster. 

It is numerically challenging to follow the entire accretion process ($\sim 10^5$ yr)
because the computational timestep becomes extremely short ($\lesssim$ yr)
to meet the Courant condition of the dynamics
of dense protostellar cores.
In order to follow the late phases of accretion and disk fragmentation, in a future study, 
we will continue the simulations by using additional numerical techniques such as sink particles, 
with which the dense region is masked and replaced with Lagrangian particles.

\subsubsection{Formation of EMP stars with peculiar abundances}
In this series of papers, we have investigated the formation processes of
C-normal and CEMP-no stars.
For the formation of other classes of EMP stars, additional nucleosynthesis mechanisms are required to 
be considered.
First, s-process elements enhancement of CEMP-s (CEMP Group I) stars can be explained by
the binary mass transfer from companion stars in the asymptotic giant branch (AGB) phase
\citep{Suda04, Komiya20}.
Since the progenitors are low-mass ($1$--$7 \ \Msun$), the companion stars 
begin to contribute to the enrichment in more massive and metal-rich halos 
($\abH{Fe} \gtrsim -3$; Fig. \ref{fig:C_Fe})
than MHs.
Therefore, to see the evolution of low-mass stars and effect of binary transfer, 
it is required to extend our simulations spatially and temporarily.

Also, rotating stars (``spinstars'') can produce lighter elements (C--Al) and trans-iron elements 
in their envelopes \citep{Meynet06, Choplin20}.
The envelopes are blown away by SN explosions and contribute to C
and s-process element enhancement \citep{Choplin19}.
Since rotating stars also produce nitrogen,
the formation of N-enhanced stars such as ${\rm CS~22952-015}$ and ${\rm BS~16550-087}$
could be explained with the spinstar models.
Some spinstars are considered to be associated with jet-like SNe \citep{Tominaga09}.
This aspherical explosion model can reproduce Fe-deficient abundance patterns
even with larger explosion energies than spherical explosion models
because Fe can fall back onto the compact remnants from the equatorial plane.
As a result, the elemental abundances of the ejecta depend on the direction from the jet axis.
\citet{Ezzeddine19} speculate that Zn-enhancement of the star ${\rm HE~1327-2326}$
could be explained by the external enrichment of a cloud 
in a specific angle range from the jet axis of a SN.
We will explore the effect of the aspherical structure of ejecta
on the metal enrichment in forthcoming papers.

\section{Conclusion}
\label{sec:conclusion}

CEMP stars, the most iron-poor class of stars, hold fossil
records of the chemical enrichment in the early Universe \citep{deBennassuti17}, and
their origin is still under debate.
In this work, we follow the metal enrichment from faint SNe
and gravitational collapse of the enriched clouds with high-resolution numerical
simulations.
We find that C abundance of the clouds is 
$\abA{C} = 3.80$, $5.06$, and $4.90$
for faint SN models with progenitor masses $\MPopIII = 13$, $50$, and $80 \ \Msun$, respectively.
These are smaller than the abundance range of observed CEMP stars ($\abA{C} > 6$)
while the abundance of a cloud enriched by a normal CCSN is consistent with the
metallicity range of observed C-normal stars \citepalias{Chiaki19}.
This behavior occurs because the mass of metals produced by faint SNe and 
the fraction of metals incorporated into the clouds are smaller than in normal CCSN model.
We also find that our detailed chemistry/cooling model enables us to 
follow the entire process of fragmentation.
AC grain cooling induces fragmentation of the enriched cloud for model {\tt F13} 
even with the lowest C abundance $\abA{C} = 3.80$.
This result indicates that low-mass stars with metallicities
$\abH{Fe} \sim -9$ may form through cooling by carbon dust grains produced by 
faint SNe and re-accreted onto the same MH.
Such ``giga metal-poor'' stars may be observed in the future as a larger number of samples of 
EMP stars will be collected.

So far $\sim 5000$ metal-poor stars have been observed.
With future observational facilities, Subaru/PFS \citep{Takada14} and 
TMT \citep{Skidmore15},
samples will increase.
To prepare for future surveys, it is urgent to study the origin of metal-poor stars from the theoretical side.
We have focused on the formation process of C-normal and CEMP-no stars in this series of papers.
The other classes of stars, which show neutron-capture element enhancement
and peculiar elemental abundances, are believed to form by multiple types of progenitors.
We will extend our numerical model to enable comprehensive studies of their formation 
processes, while being constrained by large samples of observed metal-poor stars.
Such a study should reveal elusive details of the chemical enrichment and star formation in the early Universe. 

\section*{ACKNOWLEDGMENTS}

GC is supported by Research
Fellowships of the Japan Society for the Promotion of Science (JSPS). 
JHW is supported by National Science Foundation grants AST-1614333 and
OAC-1835213, and NASA grants NNX17AG23G and 80NSSC20K0520.
The simulation was performed
with NSF's XSEDE allocation AST-120046 on the Comet and Stampede2
resources and also on the Georgia Tech PACE compute system. 
The figures in this paper are constructed with the
plotting library {\sc matplotlib} \citep{matplotlib}.




\label{lastpage}


\begin{thebibliography}{}
\bibitem[Abel et al.(2002)]{Abel02} Abel, T., Bryan, G.~L., \& Norman, M.~L.\ 2002, Science, 295, 93 
\bibitem[Aoki et al.(2007)]{Aoki07} Aoki, W., Beers, T.~C., Christlieb, N., et al.\ 2007, \apj, 655, 492 
\bibitem[Asplund et al.(2009)]{Asplund09} Asplund, M., Grevesse, N., Sauval, A.~J., \& Scott, P.\ 2009, \araa, 47, 481 
\bibitem[Audouze \& Silk(1995)]{Audouze95} Audouze, J., \& Silk, J.\ 1995, \apjl, 451, L49 



\bibitem[Beers \& Christlieb(2005)]{Beers05} Beers, T.~C., \& Christlieb, N.\ 2005, \araa, 43, 531 
\bibitem[\protect\citeauthoryear{Belczynski et al.}{2010}]{Belczynski10} Belczynski K., Bulik T., Fryer C.~L., Ruiter A., Valsecchi F., Vink J.~S., Hurley J.~R., 2010, ApJ, 714, 1217
\bibitem[Bromm et al.(1999)]{Bromm99} Bromm, V., Coppi, P.~S., \& Larson, R.~B.\ 1999, \apjl, 527, L5 
\bibitem[\protect\citeauthoryear{Bryan et al.}{1995}]{Bryan95} Bryan G.~L., Norman M.~L., Stone J.~M., Cen R., Ostriker J.~P., 1995, CoPhC, 89, 149
\bibitem[\protect\citeauthoryear{Bryan \& Norman}{1997}]{Bryan97} Bryan G.~L., Norman M.~L., 1997, arXiv, astro-ph/9710187
\bibitem[Bryan et al.(2014)]{Bryan14} Bryan, G.~L., Norman, M.~L., O'Shea, B.~W., et al.\ 2014, \apjs, 211, 19 

\bibitem[Caffau et al.(2011)]{Caffau11} Caffau, E., Bonifacio, P., Fran{\c c}ois, P., et al.\ 2011, \nat, 477, 67 
\bibitem[Cayrel et al.(2004)]{Cayrel04} Cayrel, R., Depagne, E., Spite, M., et al.\ 2004, \aap, 416, 1117 
\bibitem[Clark et al.(2011)]{Clark11} Clark, P.~C., Glover, S.~C.~O., Smith, R.~J., et al.\ 2011, Science, 331, 1040 
\bibitem[Chiaki et al.(2015)]{Chiaki15} Chiaki, G., Marassi, S., Nozawa, T., et al.\ 2015, \mnras, 446, 2659 
\bibitem[Chiaki et al.(2016)]{Chiaki16} Chiaki, G., Yoshida, N., \& Hirano, S.\ 2016, \mnras, 463, 2781 
\bibitem[Chiaki et al.(2017)]{Chiaki17} Chiaki, G., Tominaga, N., \& Nozawa, T.\ 2017, \mnras, 472, L115 
\bibitem[Chiaki et al.(2018)]{Chiaki18} Chiaki, G., Susa, H., \& Hirano, S.\ 2018, \mnras, 475, 4378
\bibitem[Chiaki \& Wise(2019)]{Chiaki19} Chiaki, G., \& Wise, J.~H.\ 2019, \mnras, 482, 3933 
\bibitem[\protect\citeauthoryear{Chon \& Hosokawa}{2019}]{Chon19} Chon S., Hosokawa T., 2019, MNRAS, 488, 2658
\bibitem[\protect\citeauthoryear{Choplin, Tominaga \& Ishigaki}{2019}]{Choplin19} Choplin A., Tominaga N., Ishigaki M.~N., 2019, A\&A, 632, A62
\bibitem[\protect\citeauthoryear{Choplin \& Hirschi}{2020}]{Choplin20} Choplin A., Hirschi R., 2020, arXiv, arXiv:2001.02341




\bibitem[de Bennassuti et al.(2014)]{deBennassuti14} de Bennassuti, M., Schneider, R., Valiante, R., \& Salvadori, S.\ 2014, \mnras, 445, 3039 
\bibitem[\protect\citeauthoryear{de Bennassuti et al.}{2015}]{deBennassuti15} de Bennassuti M., Schneider R., Valiante R., Salvadori S., 2015, MNRAS, 451, 2108
\bibitem[de Bennassuti et al.(2017)]{deBennassuti17} de Bennassuti, M., Salvadori, S., Schneider, R., Valiante, R., \& Omukai, K.\ 2017, \mnras, 465, 926 
\bibitem[Draine \& Bertoldi(1996)]{Draine96} Draine, B.~T., \& Bertoldi, F.\ 1996, \apj, 468, 269 

\bibitem[\protect\citeauthoryear{Efstathiou et al.}{1985}]{Efstathiou85} Efstathiou G., Davis M., White S.~D.~M., Frenk C.~S., 1985, ApJS, 57, 241
\bibitem[\protect\citeauthoryear{Ezzeddine et al.}{2019}]{Ezzeddine19} Ezzeddine R., et al., 2019, ApJ, 876, 97


\bibitem[\protect\citeauthoryear{Fukushima, Omukai \& Hosokawa}{2018}]{Fukushima18} Fukushima H., Omukai K., Hosokawa T., 2018, MNRAS, 473, 4754

\bibitem[\protect\citeauthoryear{Graziani et al.}{2015}]{Graziani15} Graziani L., Salvadori S., Schneider R., Kawata D., de Bennassuti M., Maselli A., 2015, MNRAS, 449, 3137
\bibitem[Greif et al.(2012)]{Greif12} Greif, T.~H., Bromm, V., Clark, P.~C., et al.\ 2012, \mnras, 424, 399 

\bibitem[Hartwig et al.(2018)]{Hartwig18} Hartwig, T., Yoshida, N., Magg, M., et al.\ 2018, \mnras, 478, 1795 
\bibitem[Hahn \& Abel(2011)]{Hahn11} Hahn, O., \& Abel, T.\ 2011, \mnras, 415, 2101 
\bibitem[Hanawa \& Matsumoto(2000)]{Hanawa00} Hanawa, T., \& Matsumoto, T.\ 2000, \pasj, 52, 241 
\bibitem[\protect\citeauthoryear{Haynes \& Kobayashi}{2019}]{Haynes19} Haynes C.~J., Kobayashi C., 2019, MNRAS, 483, 5123
\bibitem[Hirano et al.(2014)]{Hirano14} Hirano, S., Hosokawa, T., Yoshida, N., et al.\ 2014, \apj, 781, 60 
\bibitem[Hirano et al.(2015)]{Hirano15} Hirano, S., Hosokawa, T., Yoshida, N., Omukai, K., \& Yorke, H.~W.\ 2015, \mnras, 448, 568 
\bibitem[Hirano \& Bromm(2017)]{Hirano17} Hirano, S., \& Bromm, V.\ 2017, \mnras, 470, 898 
\bibitem[Hosokawa et al.(2011)]{Hosokawa11} Hosokawa, T., Omukai, K., Yoshida, N., \& Yorke, H.~W.\ 2011, Science, 334, 1250 
\bibitem[Hosokawa et al.(2016)]{Hosokawa16} Hosokawa, T., Hirano, S., Kuiper, R., et al.\ 2016, \apj, 824, 119 
\bibitem[\protect\citeauthoryear{Hunter}{2007}]{matplotlib} Hunter J.~D., 2007, CSE, 9, 90

\bibitem[\protect\citeauthoryear{Inoue \& Yoshida}{2020}]{Inoue20} Inoue S., Yoshida N., 2020, MNRAS, 491, L24
\bibitem[Ishigaki et al.(2014)]{Ishigaki14} Ishigaki, M.~N., Tominaga, N., Kobayashi, C., \& Nomoto, K.\ 2014, \apjl, 792, L32
\bibitem[Ishigaki et al.(2018)]{Ishigaki18} Ishigaki, M.~N., Tominaga, N., Kobayashi, C., \& Nomoto, K.\ 2018, arXiv:1801.07763 




\bibitem[Keller et al.(2014)]{Keller14} Keller, S.~C., Bessell, M.~S., Frebel, A., et al.\ 2014, \nat, 506, 463 
\bibitem[Kitayama \& Yoshida(2005)]{Kitayama05} Kitayama, T., \& Yoshida, N.\ 2005, \apj, 630, 675
\bibitem[\protect\citeauthoryear{Komiya et al.}{2020}]{Komiya20} Komiya Y., Suda T., Yamada S., Fujimoto M.~Y., 2020, arXiv, arXiv:2001.01420
\bibitem[Kozasa \& Hasegawa(1987)]{Kozasa87} Kozasa, T., \& Hasegawa, H.\ 1987, Progress of Theoretical Physics, 77, 1402 
\bibitem[\protect\citeauthoryear{Krumholz et al.}{2009}]{Krumholz09} Krumholz M.~R., Klein R.~I., McKee C.~F., Offner S.~S.~R., Cunningham A.~J., 2009, Sci, 323, 754

\bibitem[Lai(2000)]{Lai00} Lai, D.\ 2000, \apj, 540, 946 
\bibitem[Larson(1969)]{Larson69} Larson, R.~B.\ 1969, \mnras, 145, 271 
\bibitem[\protect\citeauthoryear{Liao, Turk \& Schive}{2019}]{Liao19} Liao W.-T., Turk M., Schive H.-Y., 2019, arXiv, arXiv:1911.07898
\bibitem[Limongi \& Chieffi(2012)]{Limongi12} Limongi, M., \& Chieffi, A. 2012, ApJS, 199, 38 




\bibitem[Marassi et al.(2014)]{Marassi14} Marassi, S., Chiaki, G., Schneider, R., et al.\ 2014, \apj, 794, 100 
\bibitem[Marassi et al.(2015)]{Marassi15} Marassi, S., Schneider, R., Limongi, M., et al.\ 2015, \mnras, 454, 4250
\bibitem[Mayer \& Duschl(2005)]{Mayer05} Mayer, M., \& Duschl, W.~J.\ 2005, \mnras, 358, 614 
\bibitem[\protect\citeauthoryear{McKee \& Ostriker}{1977}]{McKee77} McKee C.~F., Ostriker J.~P., 1977, ApJ, 218, 148
\bibitem[\protect\citeauthoryear{Meynet et al.}{2006}]{Meynet06} Meynet G., Ekstr{\"o}m S., Maeder A., 2006, A\&A, 447, 623

\bibitem[\protect\citeauthoryear{Nomoto et al.}{2006}]{Nomoto06} Nomoto K., Tominaga N., Umeda H., Kobayashi C., Maeda K., 2006, NuPhA, 777, 424
\bibitem[Nozawa et al.(2003)]{Nozawa03} Nozawa, T., Kozasa, T., Umeda, H., Maeda, K., \& Nomoto, K.\ 2003, \apj, 598, 785 
\bibitem[Nozawa et al.(2008)]{Nozawa08} Nozawa, T., Kozasa, T., Tominaga, N., et al.\ 2008, \apj, 684, 1343 

\bibitem[Omukai(2000)]{Omukai00} Omukai, K.\ 2000, \apj, 534, 809 
\bibitem[Omukai et al.(2005)]{Omukai05} Omukai, K., Tsuribe, T., Schneider, R., \& Ferrara, A.\ 2005, \apj, 626, 627 

\bibitem[Penston(1969)]{Penston69} Penston, M.~V.\ 1969, \mnras, 144, 425 
\bibitem[Planck Collaboration et al.(2016)]{Planck2015} Planck Collaboration, Ade, P.~A.~R., Aghanim, N., et al.\ 2016, \aap, 594, A13 
\bibitem[Placco et al.(2018)]{Placco18} Placco, V.~M., Beers, T.~C., Santucci, R.~M., et al.\ 2018, \aj, 155, 256 


\bibitem[Ritter et al.(2015)]{Ritter15} Ritter, J.~S., Sluder, A., Safranek-Shrader, C., Milosavljevi{\'c}, M., \& Bromm, V.\ 2015, \mnras, 451, 1190 
\bibitem[Ryan et al.(1996)]{Ryan96} Ryan, S.~G., Norris, J.~E., \& Beers, T.~C.\ 1996, \apj, 471, 254 


\bibitem[Safranek-Shrader et al.(2016)]{SafranekShrader16} Safranek-Shrader, C., Montgomery, M.~H., Milosavljevi{\'c}, M., \& Bromm, V.\ 2016, \mnras, 455, 3288 
\bibitem[Salvadori et al.(2007)]{Salvadori07} Salvadori, S., Schneider, R., \& Ferrara, A.\ 2007, \mnras, 381, 647 
\bibitem[\protect\citeauthoryear{Salvadori et al.}{2010}]{Salvadori10} Salvadori S., Ferrara A., Schneider R., Scannapieco E., Kawata D., 2010, MNRAS, 401, L5
\bibitem[Schaerer(2002)]{Schaerer02} Schaerer, D.\ 2002, \aap, 382, 28
\bibitem[\protect\citeauthoryear{Schneider et al.}{2002}]{Schneider02} Schneider R., Ferrara A., Natarajan P., Omukai K., 2002, ApJ, 571, 30
\bibitem[Schneider et al.(2003)]{Schneider03} Schneider, R., Ferrara, A., Salvaterra, R., Omukai, K., \& Bromm, V.\ 2003, \nat, 422, 869 
\bibitem[Schneider et al.(2006)]{Schneider06} Schneider, R., Omukai, K., Inoue, A.~K., \& Ferrara, A.\ 2006, \mnras, 369, 1437 
\bibitem[Schneider et al.(2012a)]{Schneider12} Schneider, R., Omukai, K., Bianchi, S., \& Valiante, R.\ 2012a, \mnras, 419, 1566 
\bibitem[Schneider et al.(2012b)]{Schneider12Caf} Schneider, R., Omukai, K., Limongi, M., et al.\ 2012b, \mnras, 423, L60 
\bibitem[\protect\citeauthoryear{Skidmore et al.}{2015}]{Skidmore15} Skidmore W., TMT International Science Development Teams, Science Advisory Committee T., 2015, RAA, 15, 1945
\bibitem[\protect\citeauthoryear{Skinner \& Wise}{2020}]{Skinner20} Skinner D., Wise J.~H., 2020, MNRAS.tmp, 126
\bibitem[Smith et al.(2008)]{Smith08} Smith, B., Sigurdsson, S., \& Abel, T.\ 2008, \mnras, 385, 1443
\bibitem[\protect\citeauthoryear{Smith et al.}{2009}]{Smith09} Smith B.~D., Turk M.~J., Sigurdsson S., O'Shea B.~W., Norman M.~L., 2009, ApJ, 691, 441
\bibitem[Smith et al.(2015)]{Smith15} Smith, B.~D., Wise, J.~H., O'Shea, B.~W., Norman, M.~L., \& Khochfar, S.\ 2015, \mnras, 452, 2822 
\bibitem[Smith et al.(2017)]{Smith17} Smith, B.~D., Bryan, G.~L., Glover, S.~C.~O., et al.\ 2017, \mnras, 466, 2217 
\bibitem[Suda et al.(2004)]{Suda04} Suda, T., Aikawa, M., Machida, M.~N., Fujimoto, M.~Y., \& Iben, I., Jr.\ 2004, \apj, 611, 476 
\bibitem[Suda et al.(2008)]{Suda08} Suda, T., Katsuta, Y., Yamada, S., et al.\ 2008, \pasj, 60, 1159 
\bibitem[Susa et al.(2014)]{Susa14} Susa, H., Hasegawa, K., \& Tominaga, N.\ 2014, \apj, 792, 32 

\bibitem[Takahashi et al.(2016)]{Takahashi16} Takahashi, S.~Z., Tsukamoto, Y., \& Inutsuka, S.\ 2016, \mnras, 458, 3597 
\bibitem[\protect\citeauthoryear{Takada et al.}{2014}]{Takada14} Takada M., et al., 2014, PASJ, 66, R1
\bibitem[Todini \& Ferrara(2001)]{Todini01} Todini, P., \& Ferrara, A.\ 2001, \mnras, 325, 726 
\bibitem[\protect\citeauthoryear{Tominaga}{2009}]{Tominaga09} Tominaga N., 2009, ApJ, 690, 526
\bibitem[Tominaga et al.(2014)]{Tominaga14} Tominaga, N., Iwamoto, N., \& Nomoto, K.\ 2014, \apj, 785, 98 
\bibitem[Truelove et al.(1997)]{Truelove97} Truelove, J.~K., Klein, R.~I., McKee, C.~F., et al.\ 1997, \apjl, 489, L179 
\bibitem[Turk et al.(2009)]{Turk09} Turk, M.~J., Abel, T., \& O'Shea, B.\ 2009, Science, 325, 601 
\bibitem[\protect\citeauthoryear{Turk et al.}{2012}]{Turk12} Turk M.~J., Oishi J.~S., Abel T., Bryan G.~L., 2012, ApJ, 745, 154
\bibitem[\protect\citeauthoryear{Turk et al.}{2011}]{yt} Turk M.~J., Smith B.~D., Oishi J.~S., Skory S., Skillman S.~W., Abel T., Norman M.~L., 2011, ApJS, 192, 9
\bibitem[Tsuribe \& Omukai(2006)]{Tsuribe06} Tsuribe, T., \& Omukai, K.\ 2006, \apjl, 642, L61 

\bibitem[Umeda \& Nomoto(2002)]{Umeda02} Umeda, H., \& Nomoto, K.\ 2002, \apj, 565, 385 
\bibitem[Umeda \& Nomoto(2003)]{Umeda03} Umeda, H., \& Nomoto, K.\ 2003, \nat, 422, 871 
 

\bibitem[\protect\citeauthoryear{Verner et al.}{1996}]{Verner96} Verner D.~A., Ferland G.~J., Korista K.~T., Yakovlev D.~G., 1996, ApJ, 465, 487


\bibitem[Wise \& Abel(2011)]{Wise11} Wise, J.~H., \& Abel, T.\ 2011, \mnras, 414, 3458 
\bibitem[Woodward \& Colella(1984)]{Woodward84} Woodward P., Colella P., 1984, J. Comput. Phys., 54, 115




\bibitem[Yoshida et al.(2003)]{Yoshida03} Yoshida, N., Abel, T., Hernquist, L., \& Sugiyama, N.\ 2003, \apj, 592, 645 
\bibitem[Yoon et al.(2016)]{Yoon16} Yoon, J., Beers, T.~C., Placco, V.~M., et al.\ 2016, \apj, 833, 20 
\bibitem[Yoon et al.(2018)]{Yoon18} Yoon, J., Beers, T.~C., Dietz, S., et al.\ 2018, arXiv:1806.04738 



\end{thebibliography}
\end{document}